\documentclass[a4paper,11pt]{article}

\usepackage{anyfontsize}
\usepackage[utf8]{inputenc}
\usepackage[english]{babel}
\usepackage{amssymb,amsmath,amsthm}
\usepackage{dsfont}
\usepackage{xcolor}
\usepackage[margin=1in]{geometry}
\usepackage{verbatim}
\usepackage{subfigure}
\usepackage{algorithm}
\usepackage[noend]{algpseudocode}
\usepackage{setspace}
\usepackage{soul}
\usepackage{changepage}
\usepackage{url}
\usepackage{natbib}
\usepackage{graphicx}%
\usepackage{authblk}

\newtheorem{assume}{Assumption}[section]

\newcommand{\supplementstart}{
	\setcounter{section}{0} 
	\renewcommand\thesection{S\arabic{section}}
	\renewcommand\theequation{S\arabic{equation}}
	\renewcommand\thefigure{S\arabic{figure}}
	\renewcommand\thetable{S\arabic{table}}
	\title{Supplementary materials for ``\maintitle"}
	\maketitle
}

\usepackage{fonttable}

\usepackage{xcolor}

\usepackage{bm}

\theoremstyle{definition}

\theoremstyle{definition}

\theoremstyle{definition}

\theoremstyle{definition}

\usepackage{mathtools}

\DeclareMathOperator{\supp}{supp}
\DeclareMathOperator{\unif}{unif}

\newcommand{\eps}{\epsilon}

\newcommand{\finepm}{PM$_{2.5}$ } 

\allowdisplaybreaks
\usepackage[shortlabels]{enumitem}

\usepackage{chngcntr}
\usepackage{apptools}
\AtAppendix{\counterwithin{lemma}{section}}

\usepackage{xr}

\usepackage[en-US]{datetime2}
\DTMlangsetup[en-US]{showdayofmonth=false}

\def\*#1{\mathbf{#1}}

\usepackage{xr-hyper}

\makeatother

\theoremstyle{plain}
\newtheorem{prop}{Proposition}

\theoremstyle{plain}
\newtheorem{thm}{Theorem}

\theoremstyle{remark}
\newtheorem{asu}{A\ignorespaces}[section]

\begin{document}
	
\newcommand{\maintitle}{A causal inference framework for spatial confounding}
\title{\maintitle}

\author[1]{Brian Gilbert}
\author[2]{Abhirup Datta}
\author[3]{Joan A. Casey}
\author[2]{Elizabeth L. Ogburn}

\affil[1]{Division of Biostatistics, NYU Grossman School of Medicine}
\affil[2]{Department of Biostatistics, Johns Hopkins Bloomberg School of Public Health}
\affil[3]{Department of Environmental \& Occupational Health Sciences, University of Washington School of Public Health}

\maketitle

\begin{abstract}
Over the past few decades, addressing ``spatial confounding" has become a major topic in spatial statistics. However, the literature has provided conflicting definitions, and many proposed solutions are tied to specific analysis models and do not address the issue of confounding as it is understood in causal inference. We offer an analysis-model-agnostic definition of spatial confounding
as the existence of an unmeasured causal confounder variable with a spatial structure. 
We present a causal inference framework for nonparametric identification of the causal effect of a continuous exposure on an outcome in the presence of spatial confounding. 
In particular, we identify two critical additional assumptions that allow the use of the spatial coordinates as a proxy for the unmeasured spatial confounder --  
 the measurability of the confounder as a function of space, which is required for conditional ignorability to hold, and the presence of a non-spatial component in the exposure, required for positivity to hold. %
 We also propose studying a causal estimand based on a ``shift intervention'' that requires less stringent identifying assumptions than traditional estimands. We then turn to estimation and focus on ``double machine learning" (DML), a procedure in which flexible models are used to regress both the exposure and outcome variables on confounders to arrive at a causal estimator with favorable robustness properties and convergence rates. This procedure avoids restrictive assumptions, such as linearity and effect homogeneity, which are typically made in spatial models and which can lead to bias when violated. We demonstrate the advantages of the DML approach analytically and via extensive simulation studies. We apply our methods and reasoning to a study of the effect of fine particulate matter exposure during pregnancy on birthweight in California.

\end{abstract}

\doublespacing
\section{Introduction}\label{sec:intro}

For most of its history, the field of geostatistics has been concerned primarily with spatial process models, where the goal is prediction and interpolation of a spatially varying outcome $Y$ \citep{banerjee,creswikle11}. However, there is now increasing interest in making inferences about the causal effect of an exposure $X$, possibly also spatially varying, on $Y$. Estimation of such causal effects needs to mitigate potential \emph{spatial confounding}: unmeasured variables that are thought to possess a spatial structure and influence both the exposure and outcome of interest.
Researchers have attempted to clarify when existing spatial models can support valid causal inferences \citep{paciorek, hodges, schnell, keller}. These conclusions have often relied on a formalism of spatial confounding that is tied to the analysis model used. A few papers have drawn from causal inference theory and methods to propose adaptations to existing spatial models \citep{papadogeorgou, schnell, guan, osama, thaden, dupont, bobb2022accounting, khan2023re}. These approaches still rely to some degree on parametric assumptions (e.g., linearity) used in the analysis model. 

We take the standpoint that spatial confounding should be defined in a manner agnostic to the analysis model used, and we
 provide a causal framework to define and study spatial confounding for geospatial (continuous or point-referenced) data. We thoroughly investigate the nonparametric causal assumptions that license the identification of causal effects of an exposure on an outcome in the presence of spatial confounding. In particular, we identify two central assumptions that allow the use of spatial coordinates as a proxy for the unmeasured spatial confounder -- i) the confounder needs to be a measurable function of space, and ii) the exposure needs to have non-spatial variation. The former is crucial to ensure conditional ignorability given the spatial coordinates. The latter is a requirement for positivity and can be viewed as a formalization of the concept of relative spatial scales of variation for the exposure and confounder, identified previously in \cite{paciorek}. We also recommend the estimation of an interpretable ``shift estimand," or the expected change in population-level outcomes if all individuals' exposures were shifted by a constant amount. Identification of the shift estimand requires a less stringent and more realistic positivity assumption than identification of the entire dose-response curve. If the outcome is linear in the exposure, this estimand corresponds to the exposure coefficient in a linear model, but it is interpretable even when the outcome-exposure relationship is non-linear.
 
In light of the proposed causal framework, we review existing methods that have been used to alleviate spatial confounding and highlight how many of them fail to address the problem or do so only under restrictive assumptions. We then turn to the application of robust methods that have been extensively studied in the causal inference literature but have, to our knowledge, not been used in the spatial literature. ``Double machine learning" (DML) methods use flexible models for regressions of both the outcome and exposure on covariates. %
We demonstrate empirically that these methods can improve the efficiency, robustness, and interoperability of inference about causal effects in spatial data. 

We begin by reviewing causal inference and confounding (Section \ref{sec:confounding}) before presenting new results on the identification of causal effects in the presence of spatial confounding (Section \ref{sec:identification}). We then review existing methods for spatial causal inference and spatial confounding, with a focus on the assumptions that must be met in order for each method to provide valid causal inferences (Section \ref{sec:models}). In Section \ref{sec:flexible}, we demonstrate how flexible machine learning methods can be used in lieu of existing methods for spatial causal inference. These are not limited by parametric assumptions such as effect linearity and homogeneity. These methods are common in the theoretical causal inference literature, but we are not aware of them having been used previously in spatial settings. We demonstrate their performance in simulations (Section \ref{sec:simulations}) and in the assessment of the effect of air pollution exposure during pregnancy on birthweight (Section \ref{sec:application}).

\section{Spatial and causal confounding}\label{sec:confounding}

The term ``spatial confounding" is frequently used to explain why point estimates from spatial models may be different from point estimates from naive models like ordinary least squares (OLS) that do not use any spatial information.  However, this term has rarely been defined explicitly, and we have identified at least four phenomena that are sometimes referred to as ``spatial confounding":

\begin{enumerate}
    \item \textbf{Omitted confounder bias}: The existence of an unmeasured confounding variable with a spatial structure and that influences both the exposure and the outcome [e.g. \cite{schnell}].
    \item \textbf{Random effect collinearity}: The change in fixed effect estimates that may come about when spatially-dependent random effects added to a regression model are collinear with the covariates [e.g. \cite{hodges}].
    \item \textbf{Regularization bias}: The finite-sample bias of methods that use flexible regression functions like splines or Gaussian processes (GP) to control for an unknown function of space [e.g. \cite{dupont}]. 
    \item \textbf{Concurvity}: The difficulty of assessing the effect of an exposure which is, or is close to, a smooth function of space, if an arbitrary smooth function of space is also included in a regression model [discussed in \cite{ramsay, paciorek}].
\end{enumerate}

The first notion of spatial confounding is consistent with standard causal definitions of confounding and is the definition that we will use throughout. The others are statistical notions tied to specific analysis models and need not be related to causal confounding. Furthermore, as we explain below, %
 correctly accounting for confounding should in general entail a change in effect estimates. 
 In the rest of this section, we will review causal inference perspectives on confounding and discuss how they apply specifically in spatial settings.

\subsection{A brief review of causal inference concepts}\label{sec:overview}

Causal effects are typically defined in terms of \emph{potential outcomes} or \emph{counterfactuals}. Let $Y$ be an outcome and $X$ an exposure of interest. A potential or counterfactual outcome $Y(x)$ is the outcome that would have been observed if, possibly contrary to fact, the exposure had been set to value $x$, holding all else constant. Typically, only one potential outcome is observed for each unit; the fact that only one of two (if $X$ is binary) or infinite (if $X$ is continuous) potential outcomes can be observed is known as the ``fundamental problem of causal inference" \citep{rubin}. The challenge, then, is to make inferences about the expectation or distribution of potential outcomes despite not being able to fully observe them.

The distribution $\mathcal{P}_{full}(\{Y(x),x\in \supp(X)\}, X, C)$, where $C$ is the set of measured covariates, is known as the \emph{full data distribution}; it is the joint distribution of covariates, exposure, and the (partially unobserved) potential outcomes. The \emph{observed data distribution} $\mathcal{P}_{obs}(Y, X, C)$ is the joint distribution of covariates, exposure, and the observed outcome. We say that a causal estimand of interest (typically a functional of the full data distribution) is \emph{identified} in a causal model if it can be written as a functional of the observed data distribution. 

We first consider the identification of the causal estimand $E[Y(x)]$. If this is identified for all $x$ then the entire dose-response curve is identified, and so are causal effects such as $E[Y(x)]-E[Y(x')]$, i.e., the expected change in potential outcomes if $X$ were set to $x$ compared to $x'$. In general, $E[Y(x)]$ is not identified from the observed data, but it can be identified with the help of measured covariates $C$ under the following assumptions: 
\begin{asu}
    [Consistency] \label{consist}  If $X_i = x$, then $Y_i= Y_i(x)$. 
\end{asu}     
\begin{asu}[Positivity]\label{posit}  For all $x$ in the support of $X$ and $c$ in the support of $C$, $(x,c)$ is in the support of $(X,C)$. 
\end{asu}
\begin{asu}[Conditional ignorability]\label{ignor} For all $x$ in the support of $X$, $Y(x) \perp X | C$.
\end{asu}

The consistency assumption implies that potential outcomes are well-defined; it would be violated if there were multiple versions of treatment $X=x$ resulting in different potential outcomes, or if one individual's treatment may affect others' outcomes (i.e., if ``interference" is possible). Positivity, also sometimes called ``overlap," ensures confounders have the same support for all possible exposure values. It is often stated in terms of conditional rather than joint probabilities: $f_{X|C}(x|c)>0$ for all $x$ in the support of $X$ and $c$ in the support of $C$, where $f_{X|C}$ is the conditional distribution of $X$ given $C$.  Confounding is present when ignorability does not hold marginally, that is $Y(x) \not\perp X$, and a covariate $C$ may be referred to as a confounder if conditioning on it helps to achieve ignorability \citep{vanderweele}. The conditional ignorability assumption is also known as the ``no unmeasured confounding" or ``conditional exchangeability" assumption. (See \cite{hernanrobins} for a more in-depth introduction to these identifying assumptions.)

 Under assumptions \ref{consist}-\ref{ignor}, $E[Y(x)]$ can be identified as follows:
 
\begin{align}
    E[Y(x)] &= E[E[Y(x) | C]] =E[E[Y(x)|X,C]]\label{g-form}\\
    &= E[ E[Y | X=x,C]] \nonumber
\end{align}
where the second equality follows from the conditional ignorability assumption, the third from the consistency assumption, and the positivity assumption ensures that the conditioning event has positive probability. The last term in (\ref{g-form}) is a functional of the observed data alone, known as the \emph{identifying functional}. %
Because this set of identifying assumptions is fully nonparametric (that is, they do not place any restrictions on the observed data distribution), we say that $E[Y(x)]$ is \emph{nonparametrically identified}.  

Estimation strategies that use the expression in (\ref{g-form}) to estimate $E[Y(x)]$ rely on a correctly specified outcome regression. An alternative identifying expression allows estimation via \emph{propensity score} estimation instead, where the propensity score is defined as $P(X=x|C)$:
\begin{align}
   E[Y(x)] & =E\left[E\left[Y(x)|C\right]\right]  =E\left[E\left[Y(x)\frac{I_{X=x}}{P(X=x|C)}|C\right]\right]\label{prop_score}\\
 & =E\left[E\left[Y\frac{I_{X=x}}{P(X=x|C)}|C\right]\right] =E\left[Y\frac{I_{X=x}}{P(X=x|C)}\right] \nonumber
\end{align}
where $I_{X=x}$ is the indicator that $X=x$. The first equality holds by the law of iterated expectations, the second by ignorability and positivity (which ensures that the denominator is non-zero), the third by consistency, and the fourth again by iterated expectations. Typically we refer to $E[E[Y|X=x,C]]$ as the identifying functional, but note that $E[E[Y|X=x,C]]= E\left[Y\frac{I_{X=x}}{P(X=x|C)}\right]$. That is, although the expressions are different, the identifying functionals in \eqref{g-form} and \eqref{prop_score} are equal to one another.

Outcome regression and propensity score regression can be combined to generate \emph{doubly robust} estimation strategies that are consistent for $E[E[Y |C,X=x]] =E\left[Y\frac{I_{X=x}}{P(X=x|C)}\right]$ if either one, but not necessarily both, of the outcome regression and propensity score models are correctly specified.  We return to doubly robust estimation in Section \ref{sec:flexible}. 

\subsection{Towards a causally informed definition of spatial confounding}\label{sec:defn}

We can now recast the first notion of spatial confounding, omitted variable bias due to a spatially varying unmeasured confounder, in more precise terms. If researchers are interested in estimating the causal effect of a (possibly spatially varying) exposure $X$ on a spatially varying outcome $Y$, then we will say that there is unmeasured spatial confounding if $Y(x) \not\perp X | C$ but $Y(x) \perp X | C,U$, where $C$ are measured confounders and $U$ is an unmeasured, spatially varying confounder. That is, ignorability does not hold conditional on measured confounders alone but it does hold with the addition of a spatially varying -- but unmeasured -- confounder $U$. This unmeasured variable might be a concrete quantity, such as noise pollution levels or household income, but in principle, it may be an abstract latent variable. 

We emphasize the following fundamental fact: in the presence of arbitrary (unstructured) unmeasured confounding, the identification and estimation of causal effects (i.e., the complete elimination of omitted variable bias) is not possible. The presence of an arbitrary unmeasured confounder leaves no imprint on the observed data, and therefore the only approach to control for unmeasured confounding or to mitigate unmeasured confounding bias is to make untestable assumptions that create a bridge between the observed data and the unmeasured confounder. 

With this fact in mind, the presence of spatial confounding may be seen as a boon contrasted with unmeasured confounding that has no spatial structure. Spatial information may be leveraged to capture some of the variability in the confounder and, possibly, to control for some of the confounding. In this case, it is the assumption of a spatial structure that creates the bridge between the observed data and the unmeasured confounder. %
In Section \ref{sec:identification}, we provide formal conditions under which spatial coordinates suffice to control for unmeasured spatial confounding, but we caution here that the mere fact that the confounder is spatially varying does not suffice to guarantee that spatial (or any) methods can mitigate or control for confounding.

\section{Identification in the presence of spatial confounding}\label{sec:identification}
 
We establish conditions for nonparametric identification of causal effects under spatial confounding using the relevant causal definitions reviewed in Section \ref{sec:overview}. 
We assume consistency (Assumption \ref{consist}) throughout, and for conciseness we omit measured covariates $C$ but note that everything that follows would also hold conditional on $C$. 
We assume the existence of a spatially varying confounder $U$ such that, had $U$ been observed, causal effects of $X$ on $Y$ would be identified. That is, we assume that ignorability and positivity hold conditional on $U$:
\begin{asu}[Ignorability conditional on unmeasured confounder $U$]\label{u-ignore}
 $Y(x) \perp X | U$,
 \end{asu}
 \begin{asu}[Positivity with respect to $U$ ]\label{u-posit} 
If $x$ is in the support of $X$ and $u$ is in the support of $U$ then  $(x, u)$ is in the support of $(X,U)$.
 \end{asu}

The question we answer in this section is: \emph{Are there assumptions under which causal effects may be estimated using spatial models even if $U$ is unobserved?} Letting $S$ denote spatial location (e.g., latitude and longitude coordinates), and assuming throughout that $S$ is observed for all units, we will present conditions under which the observed spatial location $S$ may be used as a proxy for $U$. This is the approach taken by almost all of the methods for spatial confounding considered in the literature -- they can be seen as controlling for $S$ in lieu of $U$. But as we will see, the validity of this approach depends on strong and subtle assumptions. We present one set of sufficient (nonparametric) identification assumptions; other alternatives exist (for example, the parametric propensity score model of \cite{papadogeorgou} and the approach in \cite{schnell} for the setting of area-level spatial data), but these existing alternatives are parametric and require different analytic methods. 

\subsection{Ignorability and spatial confounding}\label{ssec:ignorability}

The first crucial and strong assumption is that the unmeasured confounder $U$ is entirely spatially varying, that is all variation in $U$ is captured by $S$:%

\begin{asu}\label{ugs}
$U=g(S)$ for a fixed, measurable function $g$.
\end{asu}

Note that $U$ may not be unique; that is there may be multiple different random variables that each suffice to control for confounding. Technically, it needs only to be the case that Assumption \ref{ugs} holds for some $U$ that satisfies Assumption \ref{u-ignore}.

There are two related ways in which an arbitrary unmeasured confounder could fail to be equal to $g(S)$ for some measurable function $g$. First, $U$ may not be a function of space alone; this is the case whenever two individuals may share a location without sharing a value of $U$. For example, if $U$ is personal income but location is measured at the household level, then $U$ is not a function of $S$ alone -- though it may depend partially on $S$ -- because members of the same household may have different values of personal income.
On the other hand, even if it can be claimed that $U$ is a function of $S$, it is not necessarily a \textit{measurable} function of $S$. On a continuous domain, a measurable function can be thought of as a ``nearly continuous" function, or a continuous function with at most a small set of discontinuities (see ``Lusin's theorem" in \cite{pugh}). Then even a variable such as household income (as opposed to personal income) may not satisfy this criterion, if the map from spatial location to household income is highly discontinuous. %

If $U$ has some non-spatial variation, %
then confounding bias may be reduced by replacing $U$ with $S$, but its elimination is in general impossible. In other words, including spatial information in a causal analysis could reduce bias compared to omitting spatial information, but causal estimates would still not be consistent. 
In this setting, the flexible function of $S$ included in spatial regression models could operate like a surrogate or proxy confounder or a 
confounder measured with error, but it could not capture the confounder without error. We refer the readers to the extensive causal inference literature on this issue \citep{ogburn2012nondifferential,ogburn2013bias,kuroki2014measurement,lash2020measurement,tchetgen2020introduction,pena2021bias}. %
Hence, Assumption \ref{ugs} is indispensable for identification. Furthermore, we caution that additional assumptions are required to ensure that controlling for a proxy or mismeasured confounder mitigates, rather than exacerbates, bias \citep{ogburn2012nondifferential,ogburn2013bias,kuroki2014measurement,lash2020measurement,tchetgen2020introduction,pena2021bias}.

We make the additional assumption that conditioning on $S$ does not \emph{induce} confounding: 
\begin{asu}\label{su-ignore}
 $Y(x) \perp X | S,U$
 \end{asu}
Now we state our main proposition characterizing conditions under which conditioning on location $S$ in lieu of an unmeasured spatial confounder $U$ suffices for ignorability to hold (possibly also conditional on measured covariates $C$, omitted for conciseness).
\begin{prop}\label{prop1}
Assumptions \ref{ugs} and \ref{su-ignore} imply $Y(x) \perp X | S$.
\end{prop}

\begin{proof}
Since $U=g(S)$ is a measurable function of $S$, the double $(S,U)$ is a measurable function of $S$ as well. Since $S$ is trivially a function of $(S,U)$, there exists a measurable one-to-one mapping between $S$ and $(S,U)$; therefore $S$ and $(S,U)$ give rise to the same conditional distributions (formally, they generate the same sigma-algebra). Therefore if $Y(x) \perp X| (S,U),$ it also holds that  $Y(x) \perp X| S$.
\end{proof}

Assumption \ref{su-ignore} is similar to the conditional ignorability Assumption \ref{u-ignore}. If the latter is true, the former is often a reasonable assumption; it would typically hold whenever spatial location is causally precedent to exposure and outcome but would be violated if spatial location were a collider between exposure and outcome (i.e., it is influenced by both exposure and outcome) or a descendant of (i.e., influenced directly or indirectly by) such a collider. It would also be violated if spatial location were a mediator of the $X-Y$ relationship, e.g., if $X$ affected where people chose to live, which in turn affected $Y$. The assumption is likely uncontroversial in many applications, though this could be complicated in longitudinal studies, especially if there is movement of subjects. (See \cite{elwert} for a discussion of collider bias.) Assumption \ref{su-ignore} may be circumvented in favor of the following stronger but possibly more intuitive assumption:

 \begin{asu}\label{yxx-su-ignore}$(Y(x), X) \perp  S| U$ \end{asu}

This assumption implies that spatial location only affects the joint distribution of potential outcomes and exposure through the unmeasured confounder. This assumption is reasonable if we view $U$  as the collection of all the (unmeasured) spatial causes of $X$ and/or $Y$. Together with Assumptions \ref{u-ignore} and \ref{ugs}, this is sufficient for the conclusion of Proposition \ref{prop1}.

\subsection{Positivity and spatial confounding}\label{ssec:positivity}

A few sources \citep{weistrich, petersen, damour} have devoted themselves to discussing the details of the positivity assumption (not in the context of spatial confounding), though it receives considerably less attention in the literature than ignorability. In the context of spatial confounding,  \cite{paciorek} and \cite{schnell} associated the positivity assumption with the assertion that the exposure varies at a smaller scale than the spatial confounder, a point which is also explored in \cite{keller}, but a rigorous and general treatment of spatial-statistical concepts and the positivity assumption in causal inference is lacking in the existing literature.

As discussed in Section \ref{sec:confounding}, with outcome $Y$, exposure $X$, and a measured confounder $C$, the positivity assumption is needed to ensure that the conditioning event has positive probability in the expression $E[E[Y(x)|C, X=x]]$. If there exists $x \in \supp(X)$ such that $c \in \supp(C)$ but $(x,c) \notin \supp(X,C)$, then the above functional is not well-defined. If positivity fails, then parametric assumptions such as effect linearity or homogeneity are needed to estimate causal effects. These assumptions can be used to extrapolate from regions of the data where positivity holds to regions where it does not hold. We caution that any conclusions drawn from regions of the data where positivity does not hold are not informed by the data; they are determined entirely by the assumptions used to extrapolate. 

In the case of spatial confounding, positivity must hold with respect to $S$ in order for inference using $S$ as a proxy for $U$ to avoid extrapolation. In some settings, it may be feasible to reason about this directly, i.e., to assume 

\begin{asu}\label{s-posit}
If $x$ is in the support of $X$ and $s$ is in the support of $S$ then $(x, s)$ is in the support of $(X,S)$.
\end{asu}

Alternatively, if researchers do not have enough \textit{a priori} information about the joint distribution of exposure and spatial location to assess whether Assumption \ref{s-posit} is reasonable, positivity with respect to $S$ is also guaranteed by the following assumption in conjunction with Assumption \ref{u-posit}:

\begin{asu}\label{xsu-ignore}
$X \perp S | U$; that is, $S$ is only associated with $X$ through $U$.
\end{asu}
Note that Assumption \ref{xsu-ignore} is implied by Assumption \ref{yxx-su-ignore}. 

The following proposition says that if positivity holds conditional on $U=g(S)$, and $S$ is only associated with $X$ through $U$, then positivity also holds conditional on $S$, i.e., Assumption \ref{s-posit} follows from Assumptions \ref{u-posit} and \ref{xsu-ignore}:

\begin{prop}\label{positivity}
 Under Assumptions \ref{u-posit}, \ref{ugs} and \ref{xsu-ignore}, positivity holds conditional on $S$. \end{prop}

\begin{proof}
By the former assumption, we have $pr(X=x | U) > 0$ for each $x$. Then, by the latter assumption, $pr(X = x | U, S) > 0$. Since $(U, S) = (g(S), S)$ is a one-to-one function of $S$, conditioning on $(U,S)$ is equivalent to conditioning on $S$. Therefore, $pr(X=x | S) > 0$.
\end{proof}

In practice, the relevant consideration regarding positivity conditional on spatial location is whether there is a sufficient mix of exposure levels within reasonably small regions of the confounder space. This helps link this positivity assumption \ref{u-posit} to the condition on relative scales of variation in exposure and confounder identified by \cite{paciorek}, namely the requirement that $X$ vary at a finer spatial scale than $U$. If $X$ varies faster than $U$, then areas with similar values of $U$ will exhibit wide ranges of values of the continuous exposure $X$ and positivity is likely to hold. To our knowledge, the relative scales assumption, which is commonly cited in work on causal effects for spatial data, has not been previously formalized as a positivity condition in the causal inference framework. 

An important implication of positivity is that the exposure needs to have some source of non-spatial variation. 
If the exposure is purely a function of space, then positivity (Assumption \ref{s-posit}) is likely to be violated and the causal effect cannot be identified. We note that this requirement of a non-spatial variation in the exposure has also been used in establishing consistency of specific semi-parametric spatial models \citep{yang,dupont,gilbert2023consistency}.

\subsection{Estimability}\label{ssec:estimability}
We have shown that under Assumptions \ref{u-posit}, \ref{su-ignore}, and  \ref{xsu-ignore}; or \ref{u-ignore}, \ref{u-posit}, and \ref{yxx-su-ignore}; or \ref{su-ignore} and \ref{s-posit} (and, in any event assuming $\ref{ugs})$; causal effects are nonparametrically identified using $S$ instead of $U$: following the logic of Section \ref{sec:overview}, $E[Y(x)]=E[E[Y|X=x,S]]$. This suggests that the same estimation strategies that would be used to control for $U$ if it were observed may also be applied using $S$. Indeed, this seems to be the strategy behind most of the existing methods that we review in Section \ref{sec:models}. However, as we will discuss, many of these methods make further parametric assumptions about the functional relationships among the variables. In this section, we discuss conditions under which $U$ can be replaced by $S$ in an estimation strategy and caution that this practice may not be valid in general.

First, it must be the case that $g(S)$ is regular enough, in some sense, to be estimable by the chosen strategy. No method can approximate an arbitrarily complex function in finite samples, therefore all methods make some assumptions about the complexity or smoothness of $g(S)$. %

Second, it must be the case that the functional form of $E[Y|X=x,S]$ is compatible with the regression surface that matters for confounding, $E[Y|X=x,g(S)]$. Suppose $E[Y|X,U]$ is linear in $U$, but $U$ is a polynomial function of $S$. Then a linear regression of $Y$ onto $X$ and $S$ would be misspecified, but a polynomial regression would be correctly specified. A challenge in the spatial confounding scenario is that the relationship between $U$ and $S$ will generally be an unknown transformation, therefore it may be easier for researchers to reason about modeling assumptions on the basis of $U$ rather than $S$. This points to the need for flexible nonparametric estimation strategies that have a better chance of capturing the true confounding surface in terms of $S$.

Conversely, researchers who reason directly about the functional form of $E[Y|X,S]$ may miss important interactions or other terms needed to capture $E[Y|X,g(S)]$. Suppose the unmeasured confounder $U$ interacts with $X$ in its effect on $Y$. Then an additive model for $E[Y|X,S]$ that does not include interactions between $X$ and functions of $S$ cannot hope to control for confounding by $U$. Again, this highlights the need for flexible nonparametric estimation strategies. We discuss these issues concretely when studying various estimators in Section \ref{sec:models}.

\subsection{Shift interventions} \label{ssec:shift}

So far, we have focused on the nonparametric identification of $E[Y(x)]$,  but typically the goal of causal inference is to estimate a causal effect rather than a mean potential outcome. In the simplest case, a binary exposure $X=0,1$ admits the notion of an ``average treatment effect", defined as $E[Y(1) - Y(0)]$. When $X$ is continuous, as is often the case with environmental exposures, there is no longer just one quantity which completely specifies the average effect of interventions on $X$. One might choose to estimate the entire \textit{exposure-response curve} $E[Y(x)]$ as a function of $x$, but it can be convenient to have a more concise characterization of the effect of an intervention. 
As we will describe in Section \ref{sec:models}, this is typically accomplished in existing approaches by assuming linear and homogeneous treatment effects so that any contrast $E[Y(x)-Y(x')]$ is a multiple of the exposure coefficient in a linear regression. However, when the operative parametric assumptions do not hold, this approach does not necessarily result in interpretable causal effects. While estimation of the exposure-response curve using robust causal methods of the kind we advocate is feasible (e.g., \cite{kennedy1}), inference is not straightforward, and functional estimation may be unnecessarily complicated if a low-dimensional effect summary is sufficient. 

We define a ``shift intervention" as a special case of the class of interventions considered in \cite{haneuse2} and \cite{diaz}. A shift intervention is an intervention that assigns to individual $i$ with exposure $X_i$ the new exposure $X_i + \delta$ for some constant $\delta$ (the ``shift"). Thus the expected effect of the shift intervention on individual $i$ is $E[Y_i(X_i + \delta) - Y_i(X_i)]$; notably, this quantity depends on the observed exposure $X_i$. If we are interested in population-level effects, we may define $\Delta=E[ Y(X + \delta) - Y(X) ]$, simply taking the population average of individual effects, where the expectation is taken over the observed data distribution of $X$. The shift intervention effect $\Delta$ is the average effect of increasing the exposure of each individual in the population by $\delta$. This estimand allows the effect of the intervention to depend on the existing distribution of exposure in the population. For example, if the exposure only affects the outcome above a certain threshold, shift interventions might have no effect if current population levels are far below the threshold, but high effects if current population levels are near or above the threshold.%

This shift estimand is easily interpretable and of direct policy relevance in many applications. It is the closest nonparametric analogue to the effect captured by the exposure coefficient in a linear model (when the model is correctly specified), and estimating the shift intervention circumvents the functional estimation required to estimate the exposure-response curve.  Furthermore, while identifying the exposure-response curve without the assumption of linearity requires a very strong positivity assumption, identifying the shift effect does not.
 Recall that to estimate the full dose-response curve, the positivity requirement states that for each location $S$ in the domain, there needs to be a positive probability of observing all possible values of the exposure $X$. This may not be a realistic assumption in most applications as exposures may have limited variability over small spatial scales.

To estimate the shift estimand, we can replace the positivity assumption \ref{u-posit} with the following weaker assumption: 
\begin{asu}[Shift positivity]\label{pos-weak}
If $(x, u)$ is in the support of $(X,U)$ then $(x+\delta, u)$ is also in the support of $(X,U)$.
\end{asu} 
The \emph{shift positivity} assumption is satisfied if, whenever there is positive probability of exposure level $x$ conditional on confounder value $u$, there is also positive probability of exposure level $x+\delta$ conditional on $U=u$.
In the case of spatial confounding, if Assumption \ref{pos-weak} holds, and Assumption \ref{xsu-ignore} is satisfied, then shift positivity also holds for $S$; the logic is identical to that of the case of positivity. Alternatively, we can weaken Assumption \ref{s-posit} to a version of Assumption \ref{pos-weak} that replaces $U$ with $S$. Shift positivity may be a more reasonable assumption than positivity for exposures such as air pollution, where there may be enough variation in the exposure over relatively small regions to satisfy shift positivity, but not enough exposure variation over small regions to cover the entire range of possible exposure values.
Considering a shift intervention also requires  a weaker version of the ignorability assumption \citep{haneuse2} that only enforces ignorability for the specific exposure values under consideration:
\begin{asu}[Shift ignorability]\label{ignor-weak}
For $(x, u)$ in the support of $(X,U)$, $f(Y(x+\delta)|x,u)=f(Y(x+\delta)|x+\delta,u)$, where $f$ denotes the relevant probability density function.
\end{asu}

Under Assumption \ref{consist} and the assumptions given in Sections \ref{ssec:ignorability}, \ref{ssec:positivity} and \ref{ssec:estimability}, but with \ref{u-ignore} replaced with \ref{ignor-weak} and \ref{u-posit} replaced with \ref{pos-weak}, $E[Y(X+\delta)]$ is identified by $E[E[Y|X+\delta,S]]$, following \eqref{g-form}.

Targeting the shift estimand does not require the assumption of a linear or homogeneous model since it represents the \textit{average} expected change over all subjects if all their exposures were shifted by a certain amount, regardless of whether each subject would be expected to display the same change in outcome. The shift estimand is therefore a way of summarizing causal processes down to one scalar quantity even if the causal process is heterogeneous or non-linear. Note that when the true data-generating process is linear in $X$, the shift intervention estimand for a unit shift coincides with the ordinary linear coefficient, but the coefficient in a linear model signifies the expected change in \textit{every} subject's outcome that would be brought by a one-unit increase in exposure; for this to be meaningful both linearity and homogeneity must be satisfied. 

The shift estimand is of direct policy relevance: it is more practical to consider interventions that incrementally change each individual's exposure by some amount (for example, what would be the average effect on life-expectancy if \finepm levels were decreased everywhere by $2\mu g/ m^3$?), rather than adjusting all individuals' exposure levels to some common level $x$ which may be far from their current values. The former is captured by the shift intervention $E[Y(X)-Y(X-2)]$, where $X$ is \finepm with units $\mu g/ m^3$, while the latter is the intervention directly considered by estimation of the exposure-response curve, or by a user-specified contrast $E[Y(x)-Y(x')]$. Finally, by summarizing complex data-generating processes with a single scalar quantity, the shift estimand is valuable for its interpretability.

\section{Existing approaches to spatial confounding}\label{sec:models}

Our definition of spatial confounding and the conditions that ensure identification of the causal effect in its presence, presented in Section \ref{sec:identification}, do not rely on specific analysis models. In light of this general framework, we now assess existing methods that are commonly used to  estimate the effect of an exposure on an outcome using geospatial data.   %

\subsection{Partially linear models}\label{sec:plm}

Spatial models typically assume a data generation process with a linear effect of the exposure and  residual spatial variation in the outcome.  %
This variation, {\em the spatial effect}, can be modeled as a fixed function of space $g$ via a partially linear model (PLM)
\begin{equation}\label{eq:plm}
Y_i = \alpha + \beta X_i + g(S_i) + \epsilon_i. 
\end{equation}
Splines or other basis functions are used to model $g$. Alternatively, a linear mixed model (LMM)
\begin{equation}\label{eq:lmm}
Y_i = \alpha + \beta X_i + w(s_i) + \epsilon_i,
\end{equation}
is used with the random effects $w(s_i)$ typically modeled with Gaussian process (GP), independent of $X$, i.e., $w$ is a random function of space. The spatial effect is assumed to fully capture any spatial dependence, leaving the errors $\epsilon$ to be iid. 

 When used for spatial confounding, the (sometimes implicit) assumption is that the fixed or random function of space controls for unmeasured spatial confounding. We have identified three assumptions on the data-generating process that are commonly made in the existing spatial confounding literature:

\begin{enumerate}
    \item There exists an unmeasured variable $U$ which is a \textit{fixed} function of spatial location $S$, and $U$ influences both $X$ and $Y$. If $S$ is considered random, then $U$ is random as well, and so we may say that $U$ and $X$ are correlated. If $S$ is considered fixed, then $U$ is a nonrandom entity that is empirically associated with $X$.
    \item There exists an unmeasured variable $U$ which is a \textit{random} function of spatial location, and $U$ influences both $X$ and $Y$. In this case, $U$ is correlated with $X$ whether or not $S$ is random.
    \item There exists an unmeasured variable $U$ which is a zero-mean random function of spatial location, and $U$ influences $Y$, but not $X$. In this case, $U$ is not a ``confounder"; it merely introduces dependence among the outcomes, and naive estimators (like OLS) of the effect of $X$ on $Y$ are unbiased, although they may be inefficient or be associated with incorrect standard errors.
\end{enumerate} 

It is generally not obvious which scientific contexts justify the consideration of fixed or random effects for the locations in the data generation process. In what follows, we will mainly consider case 1; that is, we assume that there exists an unmeasured variable $U:=U(S)$, which is a fixed function of spatial location $S$ and influences both $X$ and $Y$. If the locations $S$ are considered to be sampled randomly, then $U=U(S)$ is random as well, and so we may say that $U$ and $X$ are correlated. If $S$ is considered fixed, then $U$ is a nonrandom entity that is empirically associated with $X$. Although in some cases the assumption of random sampling of locations may be unreasonable (e.g., if locations are taken on a regular lattice), the distinction between random and fixed sampling of locations may not be crucial since inference may proceed conditional on the selected locations, assuming they are ancillary to the parameters of interest. With minimal loss of generality, we will treat $S$ as random in what follows. Note that the assumption that $U$ is a fixed function of space in the true data generation process should not be confused with the modeling choice to regress $Y$ on a random function of space, e.g. using a Gaussian process regression, which may still be a valid estimation strategy (subject to assumptions) in the same way a fixed parameter can be modeled as a random quantity by assigning a prior in a Bayesian setting. 

In spatial statistics literature, it is common to model spatial effects as random functions; we see a few reasons for this. First, physical variables are rarely simple or orderly functions of space, and the complexity of real-world confounding surfaces (that is, a confounding variable as a function of two-dimensional spatial location) may be reminiscent of subjective notions of randomness. And second, Gaussian processes, a particular class of random functions, are popular for their theoretical properties, and they are empirically successful in estimating fixed functions by formally modeling them as random variables. However, there is nothing inherently \textit{random} about the true complex functions or surfaces in most applications. For example, using a Gaussian process to fit topographical elevation does not commit the researcher to any model of how that pattern of elevation came about, or to the view that topographical elevation is a random quantity. Deciding whether to model a given variable as stochastic may have more to do with generalization and targets of inference than scientific facts. If we wish to generalize to future iterations of a process that will involve a new confounding surface, then we must have a probabilistic model for this new generation. 

We caution that conceiving confounding variables as random functions of space in the true data generation process may lead to spurious claims of estimator unbiasedness. That is, in reality, a confounding variable may exist which leads to estimation error (or \textit{conditional} bias), but in imaginary re-generations of the confounding variable, the error would average out to zero (an absence of \textit{unconditional} bias). We contend that the latter fact is mostly irrelevant for scientific purposes. We illustrate this with a toy example in the supplemental Section \ref{sec:ols}. 

Conditional on the exposure $X$, the measured covariates $C$ (if any), and the locations $S$, we assume that the response $Y$ is independent and identically distributed ($iid$). Note that this assumption does not ignore spatial dependence; rather it assumes that any spatial dependence is captured via the function $U$ of the spatial locations of the observations. This is a typical assumption in many spatial regression models that include flexible functions of space in the regression function and assume that residuals are $iid$. 

For simplicity, we ignore measured confounders $C$ in this section, but all of the methods discussed below can immediately accommodate measured covariates.
In what follows in this section, we assume (\ref{eq:plm}) as the data generation model while the analysis model can be (\ref{eq:plm}) or (\ref{eq:lmm}) or other choices. %
Finally, note that some methods described below were developed in the context of areal data (spatial data where ``location" refers to a region), though in this manuscript we are most interested in geostatistical data, in which locations are point-referenced.

 \textbf{{\em Ordinary least squares (OLS):}} The naive, unadjusted estimate of the effect of $X$ on $Y$ is given by the ordinary least squares (OLS) estimator, i.e., linear regression of $Y$ on $X$. This is the baseline to which spatial confounding proposals are often compared. It is unbiased assuming the random effect model (\ref{eq:lmm}) as the true data generation process, but see Section \ref{sec:ols} of the supplement for a toy illustration of how claims of unbiasedness of the OLS estimator can be misleading. 
When data are generated from a process like (\ref{eq:plm}) with a spatial effect, it is well-known that this procedure does not adjust for spatial confounding--under any of the operative definitions thereof--and is biased in most spatial settings. See \cite{dupont} for a discussion and \cite{gilbert2023consistency} for a formal proof of the inconsistency of the OLS estimator in the context of spatial confounding. 

\textbf{{\em PLM with splines:}} The PLM makes no assumption about the relationship between $g(S)$ and $X$; if they are correlated, $g(S)$ may be interpreted as an unmeasured confounding variable. Often this assumption is embedded in the generation model for the exposure of the form
\begin{equation}\label{eq:exp}
    X_i = h(S_i) + \delta_i.
\end{equation} 
With regard to modeling $g$ in (\ref{eq:plm}) with splines, the consistency of $\hat\beta$ in PLMs is shown in \cite{rice}. However, to optimally estimate the exposure coefficient, the spline fit in a PLM must be undersmoothed. A failure to do so results in ``regularization bias," one of the many concepts that is sometimes called ``spatial confounding." A two-stage PLM procedure called {\em spatial+}, modeling both the outcome and the exposure, was proposed by \cite{dupont} in order to obviate the need to undersmooth and to alleviate regularization bias. The geoadditive structural equation model (gSEM) \citep{thaden} is a similar two-stage approach to the PLM.

\textbf{{\em Gaussian Process regression:}} The LMM analysis model (\ref{eq:lmm}) with a GP prior for the random function $w$ %
calculates probabilities under the assumption that the random effect is independent of the exposure. %
At first glance, this would seem to rule out the possibility of addressing causal confounding. However, LMMs (\ref{eq:lmm}) with GP prior can consistently estimate $\beta$ when the data are generated from a PLM  \citep{bickel2012semiparametric,yang}. Therefore both spline models and GP regression are valid analysis models for estimating the effect of exposure when the data are generated from (\ref{eq:plm}) and (\ref{eq:exp}), which accommodates scenarios of spatial confounding. 

\textbf{{\em Generalized least squares (GLS):}} GLS generalizes OLS to allow for correlated errors. GLS can also account for spatial confounding under (\ref{eq:plm}) and (\ref{eq:exp}), because it is equivalent to marginalizing out the LMM over the GP prior. The consistency of GLS under spatial confounding in the partial linear model is formally established in \citep{gilbert2023consistency}. 

\textbf{{\em Restricted spatial regression:}} %
Restricted spatial regression (RSR) \citep{hodges}, considers an analysis model that is similar to the LMM in (\ref{eq:lmm}) but restricts $w$ to be orthogonal to the exposure $X$.  %
Thus, including the spatial effects does not change the fixed effect estimate and $\boldsymbol{\hat \beta}_{RSR} = \boldsymbol{\hat\beta}_{OLS}$. 
The fact that the estimates from the mixed-effects model, like the GP regression model or the GLS estimator obtained from marginalizing out the LMM, are different from the OLS estimate was seen as a problem by \cite{hodges}. %
This overlooks the fact that OLS is inconsistent for $\beta$ in the PLM under spatial confounding, as discussed above. %
The viewpoint of \cite{hodges} can perhaps be seen as stemming from assuming the LMM (\ref{eq:lmm}), having $w$ uncorrelated with $X$, as the data-generating process, in which case the OLS is consistent. %
 In this scenario, the GLS estimate, accounting for spatial structure via the error term, will differ from the OLS estimate in finite samples, but both are unbiased estimators. %
Even assuming (\ref{eq:lmm}) as the true data-generating process,
RSR has been criticized by \cite{khan} and \cite{zimmerman} as producing anti-conservative intervals. %
However, while estimators (like GP regression and GLS)  based on (\ref{eq:lmm}) as the analysis model can be consistent under spatial confounding, as discussed above, (\ref{eq:lmm}) itself should not be considered as the data generation process as it precludes any possibility of spatial confounding by modeling $w$ to be independent of $X$. 
We emphasize that with respect to the issue of causal confounding, RSR does not even attempt to supply an unbiased effect estimate; rather, it assumes there is no confounding bias. (See \citealp{gilbert2023consistency} for further discussion of RSR, GLS, and GP regression.)

\textbf{{\em Spatially varying coefficient model (SVC):}} \cite{gelfand2003spatial} maintains the assumption of a linear outcome regression as in (\ref{eq:plm}) or (\ref{eq:lmm}) but allows the slope of the regression to vary by location, i.e., $\beta=\beta(S_i)$. This SVC model thus relaxes the assumption of effect homogeneity and can account for confounding under spatially heterogeneous linear effects. %

\subsection{Limitations of linear models}\label{ssec:limit}

Under the assumption that data are generated from a partially linear model (\ref{eq:plm}) (or an SVC) %
and the exposures are generated from
(\ref{eq:exp}), the spline- and GP-based methods above \emph{can} be used to estimate causal effects under spatial confounding. 
To see why, we note that the assumption of (\ref{eq:plm}) and  (\ref{eq:exp}) as the true data generation process implies the causal identification assumptions proposed in Section \ref{sec:identification}, and thus that controlling for a flexible function of space suffices to control for unmeasured spatial confounding. %
This is stated formally in the proposition below. The proof is provided in the Supplement.
\begin{prop}\label{ref:plm_id}
Consider the PLM (\ref{eq:plm}) with the exposure generated from (\ref{eq:exp}) where $g$ and $h$ are measurable fixed functions and $\epsilon_i$ and $\delta_i$ are i.i.d. error processes, independent of all other variables. Then there exists an $U$ for which Assumptions \ref{u-ignore}, \ref{u-posit}, \ref{ugs}, and \ref{yxx-su-ignore} are satisfied. 
\end{prop}

However, the assumption of a PLM data-generating process is stronger than all of the sets of sufficient assumptions provided in Section \ref{sec:identification} because in addition to entailing causal identification, it also entails parametric restrictions on the data-generating process. 
These parametric restrictions are homogeneous exposure effects, quantified by the coefficient $\beta$, which is assumed by all of the methods except the spatially varying coefficient model, and linear exposure effects with an additive spatial effect, which is assumed by all of the methods. These assumptions are extremely strong, and they are likely to be violated for many physical exposures. For example, common toxicological models often include a \textit{threshold}, or a value below which an exposure has negligible effect \citep{cox}, a clear violation of linearity. Similarly, air pollution may have more serious effects on older adults and children compared to younger adults \citep{gouveia}, a violation of homogeneity.

Intuitively, it may seem that departures from linearity and homogeneity ``average out" to yield an accurate estimate even in the presence of model misspecification. However, if a linear model is used to estimate a non-linear effect, the resulting estimator need not correspond to a meaningful average, as illustrated by a toy example in the Supplementary Section \ref{sec:linear}. Also,  \cite{angrist} demonstrates that the average causal effect estimated by a model which (incorrectly) assumes effect homogeneity is an average of the heterogeneous effects weighted by the conditional variance of exposure, which may be quite different from the desired uniform average. 

Recent work in spatial statistics extended the GP regression to allow for non-linear covariate
effects fitted flexibly using random forests or boosting, thereby creating GLS versions of these
machine learning procedures \citep{saha2021random,sigrist2020gaussian,zhan2024neural}. These extensions still
use additive models, implying that the effect of exposure is homogeneous in space. %
In Section \ref{sec:flexible}, we propose a class of methods that simultaneously relax both the
linearity and homogeneity assumptions.

\section{Flexible models for spatial confounding} \label{sec:flexible}

In this section, we assume that the identifying assumptions from Section \ref{sec:identification} hold; that is that the causal effect of $X$ on $Y$ is nonparametrically identified using $S$ as a proxy to control for unmeasured confounding by $U$. We now turn to estimation strategies and discuss semiparametric and nonparametric alternatives to the spatial regression models from Section \ref{sec:models}. %
We relax the standard assumption of iid regression errors and instead assume that conditional on $S$ the pairs $(Y,X)$ are identically distributed and exhibit spatial dependence according to a \emph{locally covariant random field} \citep{smith1980central}; that is, that there exists a distance $r$ such that observations sampled at any spatial locations $S_i, S_j$ such that $|S_i-S_j|>r$ are marginally independent. This holds for, e.g., spatial moving average processes and is reasonable in many applications. We also assume that the locations $S$ are sampled randomly from a geographic domain, which implies that the triples $(Y,X,S)$ exhibit only local dependence. Alternatively, under our identifying assumptions, it suffices to assume that $(Y,X,U)$ are distributed according to a locally covariant spatial random field.

\subsection{Doubly robust estimation }\label{ssec:dr}

A doubly robust estimator is consistent for the estimand of interest if at least one of two nuisance models is correctly specified, typically a model for the propensity score and a model for the outcome regression. Doubly robust versions of the nonparametric outcome regression are available, e.g., \cite{kennedy1}, but following the discussion in Section \ref{ssec:shift} we instead focus here on methods for doubly robust estimation of a shift intervention effect. Specifically, we describe the estimation of $\mu_\delta \equiv E[Y(X+\delta)]$, since the other term in the shift intervention contrast, $E[Y(X)]$, is trivially estimated by the sample average of $Y$. This estimand was first considered by \cite{diaz} and \cite{haneuse2}, with additional theory established in \cite{diaz2}.
The following provides a presentation of the estimator from \cite{diaz} for the purposes of exposition. In our simulations and data analysis, we use a specific case of the estimator presented in \cite{haneuse2} (see Supplement for details), which is asymptotically equivalent to the former, with the guarantee of the estimator lying in the space of possible values of $\mu_{\delta}$ (should the parameter space be bounded). 

Let $\hat m$ be an estimator of $E[Y | X, S]$ and $\hat \tau$ be an estimator of $f(X|S)$, conditional probability density of $X$ given $S$. Consider the estimator
\begin{multline*}
    \hat\mu_\delta = n^{-1} \sum_{i=1}^n \bigg[\frac{\hat\tau(X_i-\delta| S)}{\hat\tau(X_i|S_i)} (Y_i - \hat m(X_i, S_i))
    \\
+ \hat m(X_i+\delta, S_i) \bigg]
\end{multline*}
of $\mu_\delta \equiv E[E[Y|X+\delta, S]]$, where $\mu_\delta$ is equal to the causal parameter $E[Y(X+\delta)]$ under our identifying assumptions.

\begin{thm}\label{CLT}
Suppose that $(Y,X,S)$ are distributed identically and according to a locally covariant spatial random field. Assume an increasing domain as $n \to \infty$.  Let at least one of the following two conditions hold: (1) $\hat\tau(X,S) - f(X|S) = o_p(n^{-1/2})$ and  $\hat m(X,S) - g(X,S)= o_p(n^{-1/2})$ for some arbitrary function $g$; (2) $\hat m(X,S) - E[Y| X, S]= o_p(n^{-1/2})$ and $\hat\tau(X,S) - t(X|S)= o_p(n^{-1/2})$ for some arbitrary function $t$. Then under regularity conditions (see Supplement), $\sqrt n (\hat\mu_\delta-\mu_\delta)\to N(0,\sigma^2)$ 
with finite $\sigma^2$. %
\end{thm}

This result is doubly robust in the sense that $\hat\mu_\delta$ is consistent for $\mu_\delta$ if either one of the two nuisance functionals is consistently estimated, but not necessarily both. Asymptotic variance calculations are available for the case where one but not both of conditions (1) and (2) hold, but because in the next section we advocate using nonparametric estimation techniques for the nuisance functionals, this is not relevant for us.

\subsection{Double machine learning}

When estimating causal effects in the presence of complex or high-dimensional confounding, flexible machine learning methods provide advantages relative to more restrictive methods which risk model misspecification. However, the flexibility of these methods comes at the cost of slower rates of convergence and the resulting lack of valid inferential procedures. A solution harnesses the ``product bias" property for the doubly robust estimator proposed above:  the error in estimating $E[E[Y|X+\delta,S]]$ is the product of the errors in estimating $\tau$ and $m$ \citep{diaz2}. Then as long as the product of the rates in estimating $E[Y|X,S]$ and $f(X|S)$ is equal to $n^{-1/2}$, root-$n$ convergence can be obtained for $\mu_\delta$. This allows the use of nonparametric models without sacrificing $n^{1/2}$ rates and valid inference if, for example, both nuisance models can be estimated at rate $n^{-1/4}$. Under the standard assumption that the $Y$ is i.i.d. conditional on $X,S$ and $X$ is i.i.d. conditional on $S$,  standard nonparametric regression or machine learning methods can be used to estimate these nuisance functionals. 
A kernel smoothing estimator proposed by \cite{kennedy1} specialized this procedure for continuous exposures; the work of \cite{kennedy2} offers some general results for interventions that incrementally shift propensity scores.  
Under our relaxed spatial dependence assumption, care must be taken in using ML methods, but some recent work has shown consistency of ML methods for spatially dependent data \citep{saha2021random,goehry2020random}.

\begin{thm}\label{DML CLT}
Suppose that $(Y,X,S)$ are distributed identically and according to a locally covariant spatial random field. Assume an increasing domain as $n \to \infty$.  Let  $\hat\tau(X,S) - f(X|S) = o_p(r_n)$ and $\hat m(X,S) - E[Y| X, S]= o_p(q_n)$. Under regularity conditions (see Supplement), if $r_n \cdot q_n = o_p(n^{-1/2})$ then $\sqrt n (\hat\mu_\delta-\mu_\delta)\to N(0,\sigma^2)$ for finite $\sigma^2$.
\end{thm}

A popular approach to avoiding limiting the complexity of the models for $\tau$ and $m$ is to estimate the nuisance functionals and plug them into the estimator in independent splits of the data. The data can be split in this way many times; all of the resulting estimators are then averaged to obtain a final estimator. In our setting sample splitting is complicated by spatial dependence, but it is straightforward to use a method inspired by Bernstein blocking \citep{ibragimov1962some} to compensate for dependence. Instead of splitting the data into random subsets, we instead propose selecting a single observation at random and subsampling all of the observations within a distance of $q$ from that point, where $q$ depends on $n$ and is chosen to (approximately) achieve a target subsample size, e.g., $n/k$ for some fixed $k$. Call this set of observations $M$. We define the ``independent complement" of $M$ to be the set $M^C$ of units such that the minimum distance between any observation in $M$ and any observation in $M^C$ is at least $r$.  That is, under the assumption of a locally covariant spatial random field, $M^C$ is the set of all observations independent of $M$. Now the nuisance functionals are estimated using the data in $M$ and  $\hat\mu_\delta^{M^C}$ is calculated using the data in $M^C$. This splitting procedure can be performed many times and the resulting estimates averaged. Of course, this requires knowing the parameter $r$ governing dependence in the locally covariant spatial random field. If $r$ is unknown, researchers may still be able to provide a reasonable upper bound for $r$ and that can be used instead. Alternatively, this procedure will be asymptotically valid if the hyperparameter $r$ is allowed to depend upon and grow slowly with $n$. 

Spatial confounding is not high-dimensional in the sense of involving a high number of adjustment variables, but the complexity of nuisance functions can be expected to be considerable, since any unmeasured confounder may be a complicated function of spatial location. Therefore, DML is a strong candidate for controlling for spatial confounding.

\section{Simulations} \label{sec:simulations}

We simulate datasets involving an unmeasured spatial confounder that influences both exposure and outcome. 
The purposes of the simulations are to show the limitations of parametrically restrictive models when their assumptions do not hold, and to compare existing popular estimators with causally-informed, nonparametric DML methods. One example of this is a non-linear exposure effect where many of the methods described in Section \ref{sec:models} assuming a partially linear model becomes misspecified. We also introduce heteroskedasticity in some of the scenarios to demonstrate the potential for bias arising from the incorrect assumption of homogeneity. (Under homoskedasticity, this bias can be zero; again see \cite{angrist} which demonstrates that the homogeneous estimate is an average of the population of effects weighted by the conditional variance of exposure. Under homoskedasticity, this is equivalent to the unweighted average.)

We also simulate scenarios where no method is capable of returning consistent results due to scenarios causing a violation of the identification assumptions, such as noise in the confounding surface or smoothness of the exposure. In each case, the exposure is continuous; the causal contrast of interest is the effect of an intervention shifting the exposure by $+1$. More details of the simulation settings are provided in Supplementary Section \ref{sec:simdetails} and code can be found at
{
\url{https://github.com/bjg345/causal_spatial}
}.

The methods under consideration (with their labels in parentheses) are RSR (``RSR"), a PLM spline model (``spline"), a PLM GP model (``gp"), a spatially-varying coefficient model (``svc\_mle"), gSEM (``gSEM"), spatial+ (``spatial+"), a spline model with an interaction between exposure and location (``spline\_interaction"), a spline model with 3-dimensional splines on exposure and location (``spline3"), a random forest outcome model (``rf"), a BART outcome model (``BART"), DML with splines (``DML\_spline"), and DML with BART ('DML\_BART'). More details can be found in the Supplementary Material, Section \ref{app_sim_res}. We present the large-sample performance of these methods under scenarios involving, respectively, a non-linear effect, a structural heterogeneous effect, noisy confounding, and a smooth exposure. All other results (including those of a few methods which were only feasible in smaller samples) with additional commentary may be found in \ref{app_sim_res}.   %

\subsection{Linear spatially heterogeneous effect}
We consider the following data generation process.
\begin{equation}\label{eq:dgp1}
\begin{array}{cc}
S^a, S^b \sim_{i.i.d.} \unif(-1, 1),& 
U = \sin(2\pi S^a*S^b) + S^a + S^b\\
X \sim N(U^3, 5*\exp(U/2)),&
Y \sim N(3U+(1+U)*X, 1)
\end{array}
\end{equation}

  \begin{figure}[!t]
  \centering
  \subfigure[]{
    \includegraphics[scale=0.25]{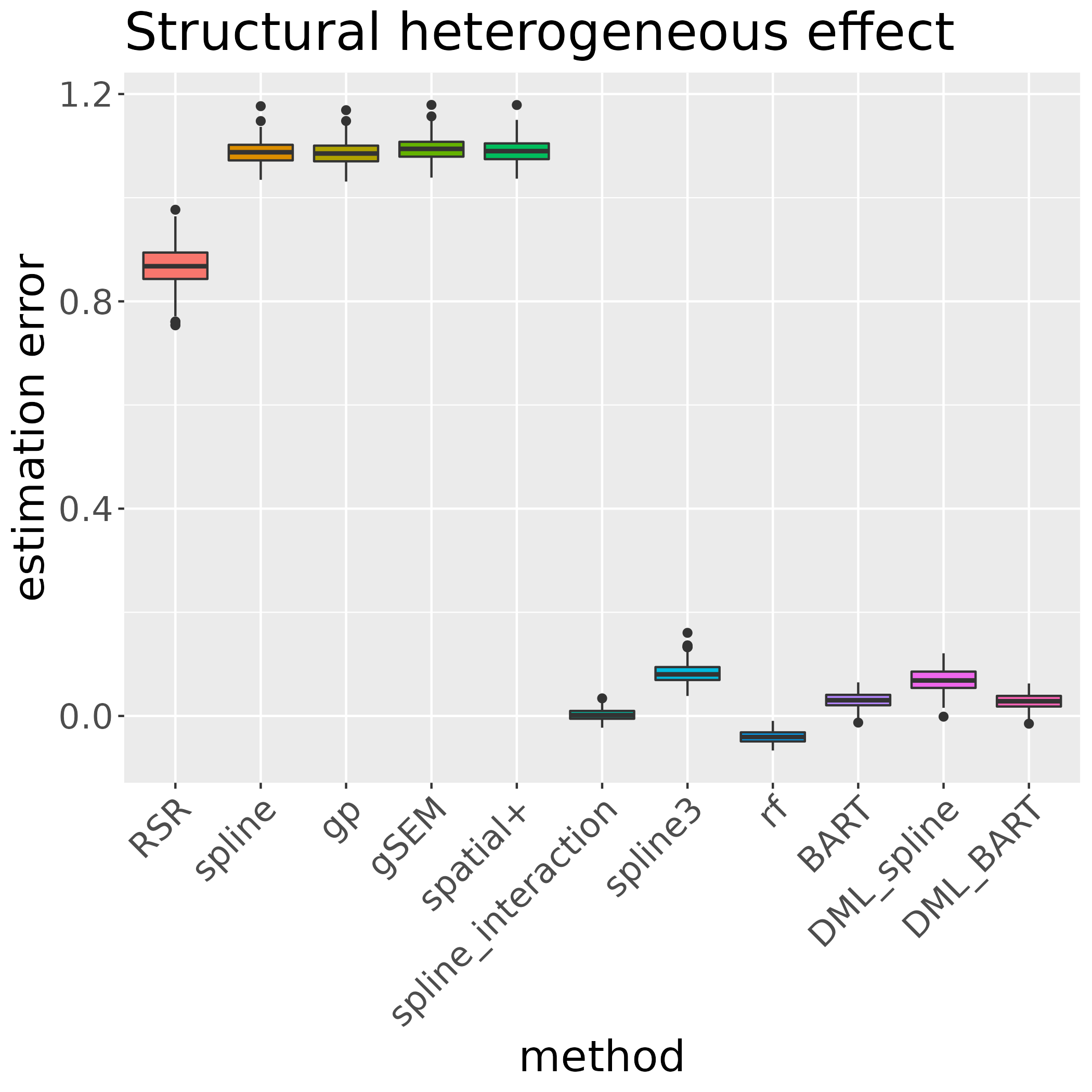}}\label{fig:sim_structhet}
    \subfigure[]{
    \includegraphics[scale=0.25]{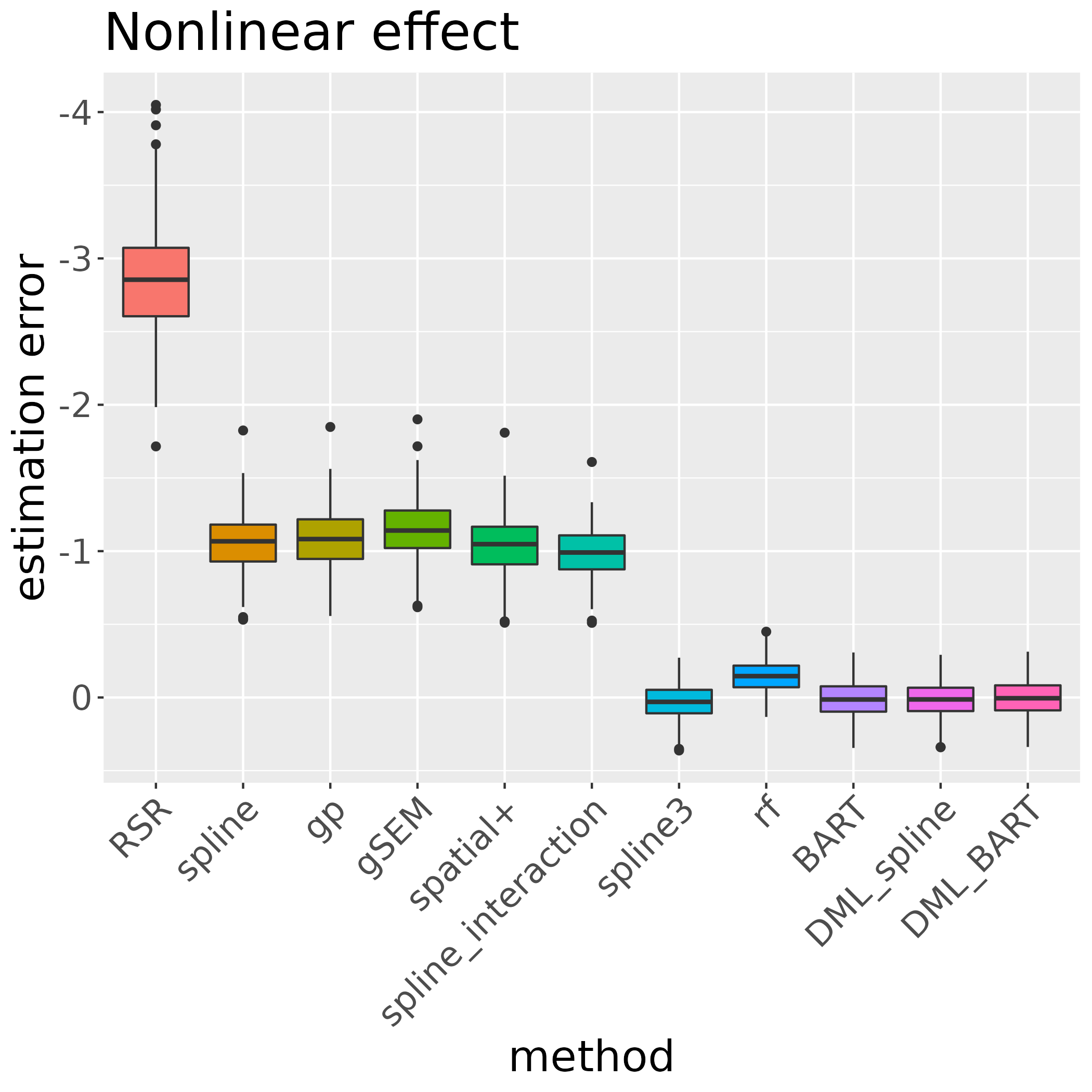}}\label{fig:sim_nonlin}\\
    \subfigure[]{\includegraphics[scale=0.25]{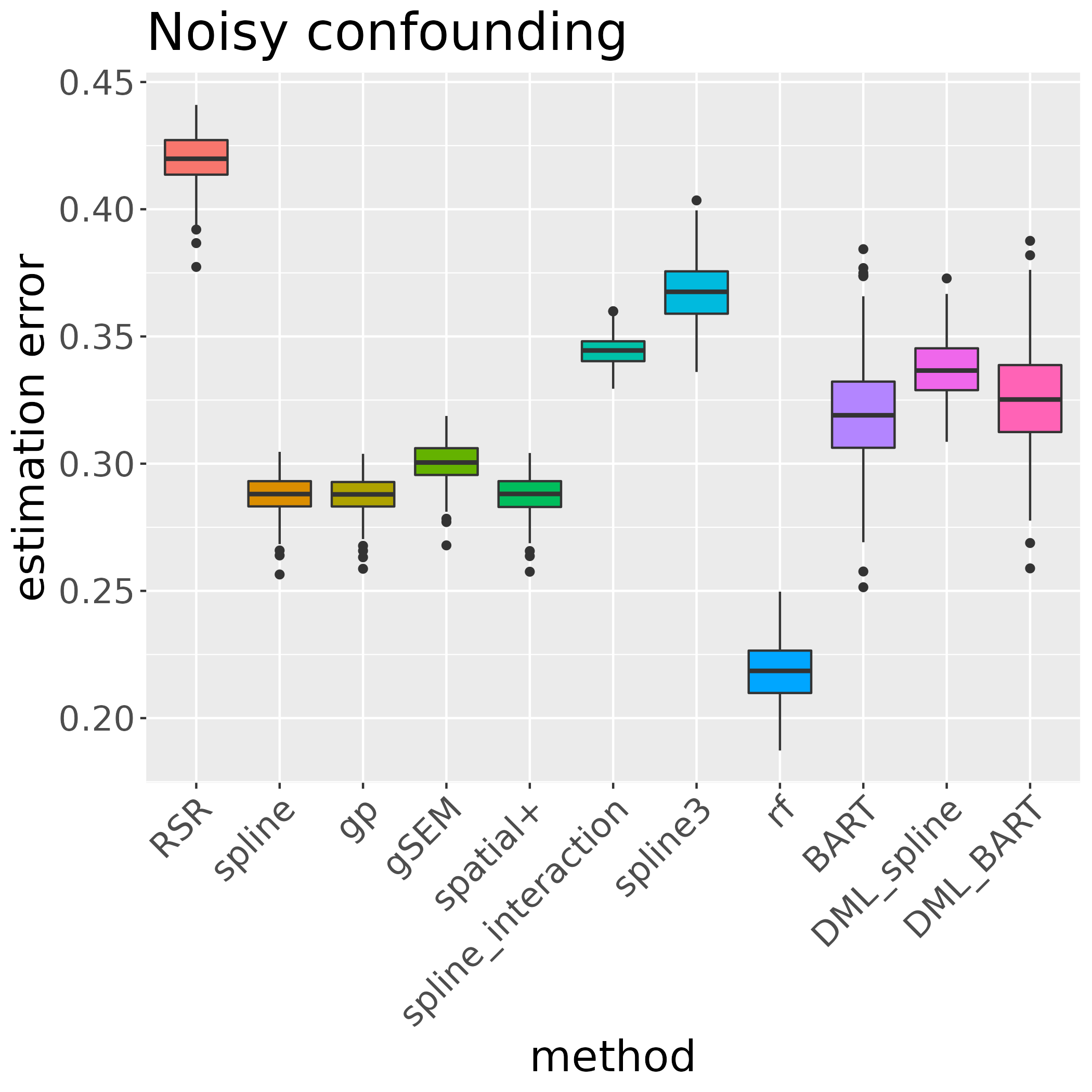}}\label{fig:sim_noisy}
    \subfigure[]{\includegraphics[scale=0.25]{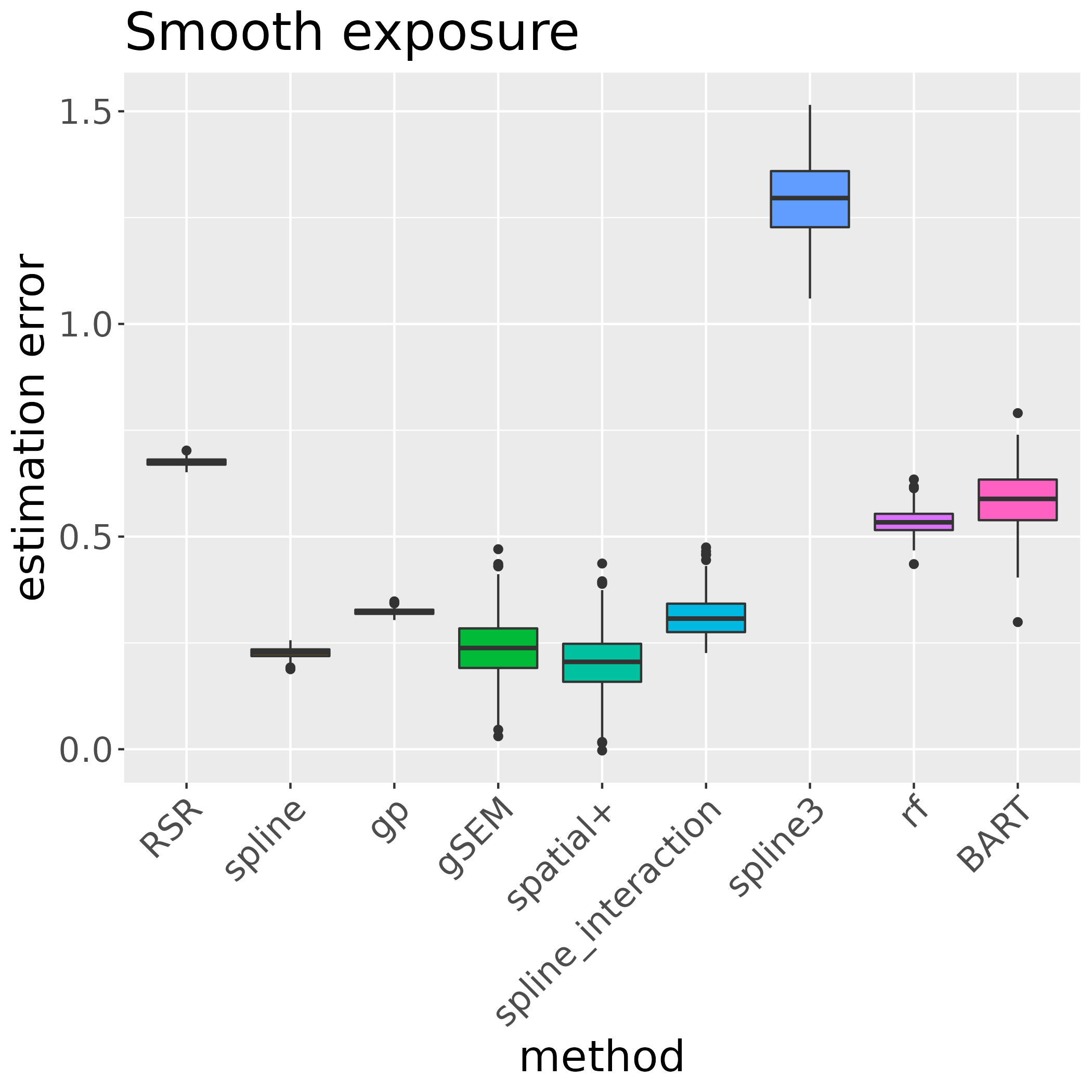}}\label{fig:sim_smooth}
    \caption{Summary of the simulation results.}
    \label{fig:sim}
\end{figure}

In this linear structural heterogeneous effect scenario (Figure \ref{fig:sim}(a)), only the flexible models and spline\_interaction are correctly specified\footnote{The exposure model of the DML methods misspecify the variances, but correctly specify the means.}. The heterogeneity, since it is driven by the confounding variable, causes severe bias in all methods that assume a homogeneous effect.

\subsection{Non-linear effect}
We now consider a dataset with non-linear but homogeneous exposure effect generated as 
\begin{align*}
X \sim N(U^3, 5),\, Y \sim N(3U+X+X^2, 1),
\end{align*}
with other specifications remaining the same. 
Only the flexible models are correctly specified in the non-linear effect scenario, Figure \ref{fig:sim}(b); all other methods exhibit substantial bias.

\subsection{Violation of the identification assumptions}

Finally, we consider scenarios where either of the two central assumptions for identification has been violated -- the assumption of the confounder being a measurable function of space (Assumption \ref{ugs}) and the positivity assumption (Assumption \ref{s-posit}) which translates to the exposure having non-spatial variation. In one scenario we use a noisy confounder and in another one, we use a spatially-smooth exposure. Details of the data generation mechanism are in Sections \ref{sec:noisy} and \ref{sec:smooth} of the supplement. We see in Figure \ref{fig:sim} (c) and (d) that under both of these scenarios, all models exhibit substantial error highlighting that all of the methods fail in situations where one of these assumptions is violated.

\section{Application }\label{sec:application}

Although the nature of the causal pathway is debated, birthweight has been widely studied for its strong correlation with health and economic outcomes in childhood and adulthood \citep{wilcox, behrman, chye}; low birthweight is a risk factor for adverse outcomes including mortality. In addition, evidence suggests that exposure to ambient \finepm during pregnancy is linked to reductions in birthweight \citep{li, basu}. A 2021 meta-analysis gave an estimated $-27.55$ $g$ [95\% CI: (-48.45, -6.65) ] change in birthweight for each $10 \mu g/m^3$ additional concentration of exposure to \finepm during pregnancy, noting a wide range of previous effect estimates in the literature, all of which appear to derive from linear models \citep{uwak}. 

In order to estimate the causal effect of ambient \finepm on birthweight, it is necessary to adjust for confounding variables. ``Greenspace" is a term with multiple definitions but generally refers to vegetation and natural spaces in residential environments \citep{taylor}. The positive association between greenspace and birthweight has been studied \citep{toda, cusack}, as has the potential mitigation of \finepm by greenspace \citep{chen}. Therefore, greenspace is a potential confounder of the relationship between \finepm exposure and birthweight. Since greenspace is a relatively smooth function of spatial location (i.e., nearby locations have similar levels of greenspace), we may attempt to control for greenspace by controlling for spatial location.

For the present study, we use records of births from the California Department of Public Health's Birth Statistical Master Files with mother's home addresses geocoded using ArcGIS; we restrict our analysis to conceptions in the year 2013. We obtain daily \finepm levels at 1-kilometer resolution from the NASA Socioeconomic Data and Applications Center \citep{sedac} and average them over the course of pregnancy. To measure greenspace, we use peak normalized difference vegetation index (NDVI) measured at 250-m resolution and aggregated at the census block level from the year 2013 \citep{didan2015modis}. In all our analyses, we include the potential confounding variables: race/ethnicity, parity, Kotelchuck index, maternal age, maternal education, and census tract-level per capita income from the 2010 American Community Survey (sourced from \cite {manson2023ipums}). We exclude individuals with missing data, as well as pregnancies with recorded gestational lengths less than or equal to 22 weeks or greater than or equal to 44 weeks, mothers with recorded ages less than 15 or greater than 49, and birthweights recorded below 500 grams. This leaves us with $n \approx 450,000$ live births. Latitude and longitude were recorded in the Web Mercator projection, and all quantitative variables were normalized prior to model-fitting.

\begin{figure}
    \centering
    \includegraphics[scale=.175]{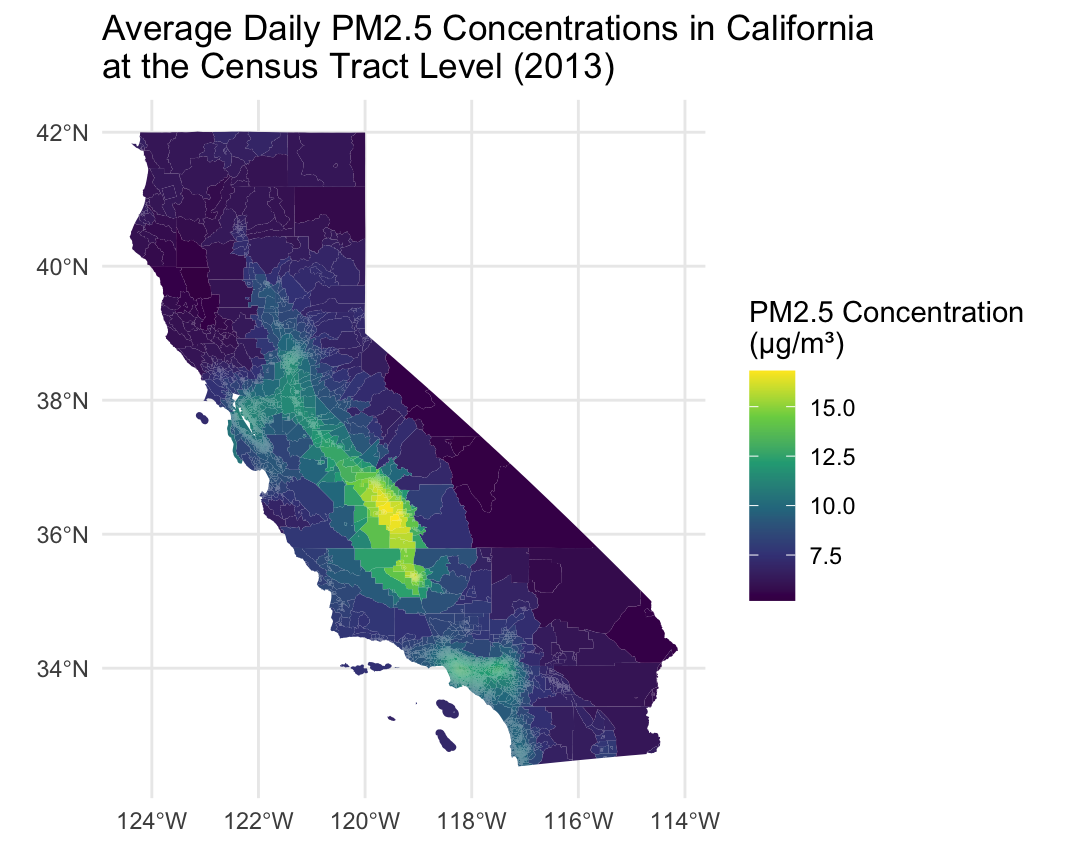}
    \caption[Average daily \finepm concentrations by census tract in 2013 (data from \cite{cdc_pm25_2011} and \cite{us_census_2010}). Note that the exposure values used in our study were more granular, as described in the text.]{PM$_{2.5}$ exposure concentrations by census tract in 2013 (data from \cite{cdc_pm25_2011} and \cite{us_census_2010}). Note that the exposure values used in our study were more granular, as described in the text.}
    \label{fig:pm25}
\end{figure}

We fit four families of models: (1) a standard PLM as in equation (\ref{eq:plm}), (2) a DML estimator in which both exposure and spatial location enter the model through flexible functions while other covariates enter the model linearly, (3) a DML estimator in which BART is used to estimate the exposure and outcome models, and (4) a modification of (3) in which spatial locations are replaced with the value of the estimated spatial smooth from outcome model in (2). We fit each model with and without each of  NDVI and spatial location. 

 The exposure shift of interest is 0.1 $\mu g/m^3$. We choose this small shift due to a lack of positivity for larger shifts. Under linearity, this effect can be multiplied by 10 in order to recover the implied average effect of a 0.1 $\mu g/m^3$ shift in exposure, and if the effect of exposure is approximately locally linear, this rescaling is likely to provide a reasonable approximation. Because 0.1 $\mu g/m^3$ is too small of an effect to be of policy relevance we recommend rescaling. After rescaling the values we obtain may be compared to the results of linear models (by appropriate rescaling), but the effective interpretation of the coefficients of those linear models relies on the assumption of effect linearity which we do not make for non-linear models.
 Except for the linear models, 95\% confidence intervals are by bootstrap standard errors. Our estimates are recorded in Table \ref{tab:ests}.

\begin{table*}[ht]
\centering
\begin{tabular}{lll}
  \hline
Method & Adjustment & Estimate \\ 
  \hline
PLM & Neither & -0.14 (-0.18, -0.09) \\ 
  PLM & NDVI & -0.11 (-0.15, -0.06) \\ 
  PLM & Spatial location & -0.10 (-0.17, -0.03) \\ 
  PLM & Both & -0.09 (-0.16, -0.02) \\
  \hline
  DML & Neither & -0.42 (-0.49, -0.34) \\ 
  DML & NDVI & -0.35 (-0.43, -0.28) \\ 
  DML & Spatial location & -0.14 (-0.24, -0.04) \\ 
  DML & Both & -0.15 (-0.26, -0.05) \\ 
  \hline
  BART DML & Neither & -0.43 (-0.51, -0.34) \\ 
  BART DML & NDVI & -0.35 (-0.45, -0.25) \\ 
  BART - smooth DML & Spatial location* & -0.30 (-0.42, -0.18) \\ 
  BART - smooth DML & Both* & -0.16 (-0.28, -0.03) \\ 
   \hline
\end{tabular}
\caption{Estimated effects of a $0.1 \mu g/m^3$ shift in average daily \finepm exposure on birthweight in grams. An asterisk indicates that the estimated smooth from the GAM model was used for adjustment in place of directly using the spatial location.}
\label{tab:ests}
\end{table*}

Beginning with the linear effect models, we see that the effect of adjusting for NDVI is similar to the effect of adjusting for spatial location. This is consistent with the understanding that spatial location stands in for the unmeasured confounder NDVI. However, an assumption of linear effect may not be consistent with air pollution exposure (see, for example, \cite{lin2022long} and \cite{miller2024nonlinear}). As we can see from Figure \ref{fig:pm25}, there appears to be a threshold effect where an effect only appears at high levels of exposure. The non-linear effect GAM models assume a homogeneous effect but do not assume the effect is linear. The estimates from these models are different from those of the linear effect models, but they display a similar pattern in which adjusting for spatial location and NDVI has a qualitatively similar effect. However, adjusting for spatial location decreases the magnitude of the estimated effect by more than adjusting for NDVI, suggesting that adjusting for space ``picks up" additional confounding variables.

We also fit BART models which assume neither homogeneity or linearity. The non-spatial estimates (Adjustment: Neither or NDVI) are very similar to the non-linear GAM estimates, but the models which adjust for space (not shown) were aberrant, suggesting a much \textit{larger} effect magnitude. We believe that in this scenario, the BART model was incapable of learning the appropriate adjustment for space; perhaps the inclusion of a larger set of covariates makes this model fitting more difficult. In addition, the large sample size makes posterior sampling computationally difficult, and the estimates may be prone to Monte Carlo error. The BART - smooth model is our attempt at a compromise, in which a BART model is fit using the estimated functional smooth from the non-linear/GAM spatial outcome model rather than spatial location itself. That is, while BART is incapable of adjusting for the raw spatial location, it may yield reasonable estimates if fed the smooth term estimated from the GAM model. (We write the adjustment sets for these models with an asterisk to denote this distinction.) We do in fact observe the BART - smooth estimates to be similar to the non-linear/GAM estimates.

We believe the most reliable estimates are those that come from non-linear models while controlling for all known covariates. These estimates are $-0.15\: (-0.26, -0.05)$ from the non-linear/GAM model and $-0.16\: (-0.28, -0.03)$ from the BART - smooth model. These estimates are concordant with each other as well as the confidence interval given by the \cite{uwak} meta-analysis of $(-48.45, -6.65)$ for a $10 \: \mu g/m^3$ increase.

Given that we fit multiple models, these results should be seen as exploratory rather than as a definite contribution to the scientific literature. Other limitations include the potential impact of ``live-birth bias" (the bias arising from only including data from live births, analogous to the ``birthweight paradox" concerning conditioning on low birth weight) \citep{goin, hernandez}, and the mismatch in measurement resolution of various covariates, including NDVI and \finepm.

 \section{Discussion}\label{sec:discussion}
 
  In this paper, we have argued that spatial confounding can be understood most clearly as omitted variable bias. We have given conditions sufficient for the identification and estimation of causal effects when unmeasured confounders vary spatially, and we have given novel results for causal estimation in the presence of both spatial confounding and spatial correlation. An important assumption is that there should not be non-spatial variation in the unmeasured confounder; if there is, then the methods described in this paper may mitigate, but would not be expected to fully control for, confounding bias. We recommend the use of flexible models to estimate causal effects rather than the restrictive (partially) linear models that are typically used in spatial statistics; the assumptions of linearity and homogeneity are often implausible in practice and can lead to substantial bias. DML methods should be considered strong candidates for spatial models, though our data analysis shows this application should be handled with care.
 
 There remains much room for further research. For example, unmeasured confounders will often be associated with spatial location even if they are not measurable functions of spatial location. In this scenario, it would be beneficial to quantify the extent to which confounding bias can be reduced by including spatial location in the causal model and/or to provide a framework for sensitivity analysis for causal estimates. The assessment of exposures that are perfectly smooth in space (for example, distance to some location such as a factory) is beyond the reach of our current methods due to violation of positivity but may be possible under certain assumptions about relevant scales of variation, as in \cite{paciorek} and \cite{keller}. In addition, the mitigation of spatial confounding for areal data has been mostly ignored in this work as some fundamental concepts are distinct in the areal domain; with discrete locations, it is not possible for a confounding variable to be a smooth function, and it is not obvious how to consider an asymptotic regime as the number of areal units is typically fixed.

Besides spatial confounding, preferential sampling of locations can be another issue which leads to distorted causal effect estimates.  \cite{diggle2010geostatistical} showed that preferential sampling can bias estimates of spatial structure (variograms) and predictions, but did not study its impact on exposure effect estimates. \cite{pati2011bayesian} considered exposures as well and showed that by modeling the intensity process for sampling of the locations, one can correct for the bias and identify the exposure effect estimate. %
 In Section \ref{sec:identification}, we present a sufficient set of assumptions under which non-parametric identification (and thereby estimation) of the exposure effect is feasible by using the locations as a proxy for the unmeasured confounder. The framework subsumes certain types of preferential sampling, e.g., ones where the intensity process for sampling depends only on the function $g$ used to define the unmeasured confounder (Assumption \ref{ugs}), but not on $X$ or $Y$. %
 However, if the sampling intensity directly depends on $X$ or $Y$, then Assumption (\ref{yxx-su-ignore}) is violated, and causal identification is no longer feasible. Thus, preferential sampling can sometimes lead to a violation of the assumptions needed for causal identification under spatial confounding, and different adjustment is required. The approaches to adjust for preferential sampling proposed by \cite{diggle2010geostatistical,pati2011bayesian} rely on correct specification of their parametric models, both for the outcome and for the sampling of the locations. Generally, spatial confounding can occur with or without preferential sampling, and vice versa. In the future, more non-parametric framing of preferential sampling and its interplay with unmeasured spatial confounding needs to be studied.

\section{Acknowledgments}
This work is partially supported by National Institute of Environmental Health Sciences (NIEHS) grant R01ES033739 and Office of Naval Research (ONR) grant  N000142112820. The authors would like to thank Heather McBrien for assistance with statistical programming. The authors are also grateful for the use of the facilities at the Joint High Performance Computing Exchange (JHPCE) in the Department of Biostatistics, Johns Hopkins Bloomberg School of
Public Health that have contributed to the research results reported within this paper.

\clearpage 
\supplementstart

\section{Proof of Proposition 1}\label{sec:prop1}

Consider the  variable $U=(g(S),h(S))$. As $g$ and $h$ are measurable functions, then $U$ is a measurable function of $S$ and Assumption \ref{ugs} is satisfied. We can write the potential outcome and exposure models as $Y(x)=x\beta + f_1(U) + \eps_i$, $X = f_2(U) + \delta_i$ where $f_i : \mathbb R^2 \to \mathbb R$ is the projection function mapping any $(v_1,v_2)$ to $v_i$ for $i=1,2$. As the error processes of each of the two models are independent of each other, conditional ignorability (Assumption \ref{u-ignore}) is met. The exposure model  with real-valued random errors ensures positivity with respect to $U$, proving Assumption \ref{u-posit}. 
Finally, we also see that $S$ enters both the potential outcome and exposure models only via $U$ thereby establishing Assumption \ref{yxx-su-ignore}.

\section{Proof of Theorems 1 and 2}\label{sec:proof}

The proof of Theorems 1 and 2 draws upon several existing literatures. \cite{diaz2} established the pathwise differentiability of the parameter $\mu_\delta$. \cite{kennedy2022semiparametric} reviews standard iid methods for proving the asymptotic normality of DML estimators. \cite{van2014causal} and \cite{ogburncausal} established asymptotic results for such estimators in the presence of (social network) dependence, and \cite{smith1980central} proved a version of the spatial central limit theorem that we invoke in the last step of the proof.

We first expand $\hat\mu_\delta-\mu_\delta$ via a von Mises expansion, letting $\mathcal{P}$ denote the true data-generating distribution, $\mathcal{P}_n$ be the empirical distribution, and $\hat{\mathcal{P}}$ be the limit of the estimated data-generating process (where what comes next will only depend on the estimated data-generating process through the estimated functionals $\tau$ and $m$). We will let $\mu_\delta(\cdot)$ denote the estimator functional evaluated at the true, empirical, or estimated distribution, respectively.

The influence function for the estimator $\hat\mu_\delta$ is given by 
\[
\psi(Y,X,S)= \frac{\tau(X-\delta| S)}{\tau(X|S)} (Y -  m(X, S)) +  m(X+\delta, S) - \mu_\delta(\mathcal{P})
\]

First, note that (following \cite{kennedy2022semiparametric})
\begin{align}
	\hat\mu_\delta-\mu_\delta & = (\mathcal{P}_n - \mathcal{P}) \{\psi(Y,X,S;\mathcal{P} )\} + 
	(\mathcal{P}_n - \mathcal{P}) \{\psi(Y,X,S;\hat{\mathcal{P}}) - \psi(Y,X,S;\mathcal{P} )\} 
	+ R_2(\mathcal{\hat P},\mathcal{P}).
\end{align}

We will prove that the first term dominates and is amenable to a spatial central limit theorem below. First, we provide conditions under which the second term is amenable to empirical process results and is therefore negligible and analyze the third term, i.e., the second-order remainder. 

\begin{assume}\label{L2 norm}
	$\psi(Y,X,S; \mathcal{\hat P})$ converges to $\psi(Y,X,S; \mathcal{ P})$ in $L_2$ norm.
\end{assume}

\begin{assume}\label{bounded entropy}
	There exists some $\eta>0$ such that $\int_0^\eta\sqrt{log(N(\epsilon,\mathcal{P},d))}d\epsilon < \infty$, where $N(\epsilon,\mathcal{P},d)$ is the number of balls of size $\epsilon$ with respect to metric $d$ required to cover $\mathcal{P}$.
\end{assume}

Assumption \ref{bounded entropy} ensures that the nuisance functionals are Donsker and is not required if sample splitting is used. 

Under these assumptions, the second term in the expansion is $o_{\mathcal{P}}(n^{-1/2})$ and does not contribute to the asymptotic distribution. This is a standard result for i.i.d. data, for dependent data it was first established by \cite{van2014causal}. Although that paper dealt with social network data, the assumption of a locally covariant spatial random field with increasing domain asymptotics satisfies the assumptions of \cite{van2014causal}. 

Now turning to the third term, it can be shown that 
\[
R_2(\mathcal{\hat P},\mathcal{P}) = \int  \left[ \frac{\hat\tau(X-\delta,S)} {\hat\tau(X,S)} 
-  \frac{\tau(X-\delta, S)}{\tau(X,S)} \right] \left[ \hat m(X,S)-m(X,S) \right] \tau(X,S) d\mathcal{P}
\]

This must also be  $o_{\mathcal{P}}(n^{-1/2})$; this will be the case as long as the product of the rates of convergence of $\hat\tau$ to $\tau$ and $\hat m$ to $m$ is $n^{-1/2}$. This covers Theorem 1, where one nuisance estimator fails to converge (i.e., has rate $O_p(1)$) while the other converges with rate $o_p(n^{-1/2})$, in addition to Theorem 2. 

Now we turn to the first, dominant term. By definition it is equal to $\Sigma_{i=1}^n \psi(Y_i,X_i,S_i) - E[ \psi(Y,X,S) ]$,  so it suffices to cite a central limit theorem for data with the dependence structure assumed for $(Y,X,S)$. The assumption of a locally covariant spatial random field is statistically equivalent to the dependency neighborhoods invoked by \cite{ogburncausal}, and the assumption of increasing domain asymptotics (which ensures that the number of observations $j$ within a distance of $r$ of any unit $i$ does not increase with $n$) implies that the size of the dependency neighborhoods is uniformly bounded as $n$ goes to infinity. Therefore, the proof follows by reference to a special case of the result in \cite{ogburncausal}.

Note that if we assumed a dependence structure for $Y,X,U$ rather than $Y,X,S$, we would be licensed to replace $U$ with $g(S)$, by assumption, and $g(S)$ with $S$ because the error in estimating $g(S)$ with $S$ was dealt with by the remainder terms.

\section{Doubly robust shift estimator}\label{app_hr}

We describe the version of a doubly robust estimator that we implemented, following \cite{haneuse2}. Let $\hat f$ be an estimator of $E[Y | X,S]$ and $\hat \tau$ be an estimator of $p(X |S)$. We solve the following estimating equation for $\gamma$:
\begin{gather*}
	\sum_{i=1}^n \hat \lambda(X_i,S_i) (Y_i - \hat f(X_i,S_i) - \gamma \hat \lambda(X_i,S_i) ) = 0
\end{gather*}
where $\hat\lambda$ means an estimate of $\lambda(X,S):= \frac{pr(X-\delta |S)}{pr(X|S)}$ based on $\hat \tau$. This equation may be solved by fitting a linear model for $Y$ with the covariate $\hat \lambda(X_i, S_i)$, no intercept, and the offset $f(X_i, S_i)$. Then the doubly robust estimate of $\mu_\delta$ is given by
\begin{gather*}
	\hat\mu_{\delta,DR} = n^{-1}\sum_{i=1}^n (\hat f(X_i+\delta, S_i) + \hat \gamma \hat\lambda(X_i+\delta, S_i))
\end{gather*}
The details of the convergence of $\hat\mu_{\delta,DR}$ can be found in the appendix of \cite{haneuse2}. 
To see the double-robustness, first assume the exposure model is consistent, without any assumption on the outcome regression. Note that for any function $h$, we have $E[ \lambda(X, S) h(X,S) ] = E[h(X+\delta, S)]$. Therefore, in large samples, the solution of the estimating equation will converge to the solution of the equation 
\begin{gather*}
	\sum_{i=1}^n \lambda(X_i, S_i) Y_i =\sum_{i=1}^n \hat f(X_i+\delta,S_i) + \gamma \lambda(X_i+\delta, S_i)
\end{gather*}
Since $E[\lambda(X_i, S_i) Y_i] = \mu_\delta$, we can divide the above equation by $n$ to see that the law of large numbers implies that $\hat \mu_{\delta, DR} \to \mu_\delta$ as well. 

On the other hand, if the outcome regression is consistent, then in large samples, the solution of the estimating equation will converge to the solution of the equation
\begin{gather*}
	\sum_{i=1}^n \hat\lambda(X_i, S_i) (Y_i - E[Y_i | X_i, S_i] - \gamma \hat\lambda(X_i, S_i)) = 0 \\
	\sum_{i=1}^n \hat\lambda(X_i, S_i) (Y_i - E[Y_i | X_i, S_i]) - \gamma \hat\lambda^2(X_i, S_i)
	\implies \\
	\gamma \to \frac{ \sum_{i=1}^n \hat\lambda(X_i, S_i) (Y_i - E[Y_i | X_i, S_i])}{\sum_{i=1}^n\hat \lambda^2(X_i, S_i)}\\
	=\frac{n^{-1} \sum_{i=1}^n \hat\lambda(X_i, S_i) (Y_i - E[Y_i | X_i, S_i])}{n^{-1}\sum_{i=1}^n\hat \lambda^2(X_i, S_i)} \\
	\to \frac{E[ \hat\lambda(X_i, S_i) (Y_i - E[Y_i | X_i, S_i])]}{E[\hat \lambda^2(X_i, S_i)]} 
\end{gather*}
Note that $Y-E[Y | X,S]$ is uncorrelated with any function of $(X,S)$. Therefore, as long as $\hat\lambda$ converges some function other than the constant 0, we have $\gamma \to 0$. Thus $\hat\mu_{\delta, DR} $ will asymptotically be based (to the first order) only on $\hat f$ and hence consistent.
When the outcome regression and exposure models are both consistent $\hat \mu_{\delta,DR}$ will achieve the highest possible asymptotic efficiency, as noted by \cite{kennedy2}.

\section{Spurious unbiasedness of OLS}\label{sec:ols}

We illustrate why conceiving of effects as ``random" may lead to spurious claims of estimator unbiasedness. In fact, any model can be expanded in such a way as to render the unadjusted estimator ``unbiased;" researchers should think carefully about whether this is in fact the setting about which they would like to make inferences. We believe it typically is not, and that spatial confounding should be viewed as a fixed, not random, effect.  Consider the following hierarchical model:
\begin{gather*}
	r \sim \mbox{Unif}(-1, 1), 
	X_i, U_i | r,\, \sim_{i.i.d.} N( \*0, \begin{bmatrix}
		1 & r \\
		r & 1 \\
	\end{bmatrix}),\,
	Y_i | X_i, U_i,r \sim_{i.i.d.} N(X_i+U_i, 1)
\end{gather*}
In any given realization of such a data set, $U$ is a confounding variable and should be included in a regression of $Y$ on $X$ to avoid a (conditionally) biased estimate of the true slope. However, in repeated sampling, as $E(r)=0$, the average correlation of $U$ and $X$ is 0, and thus the unadjusted estimator is (unconditionally) unbiased. A similar hierarchical structure can be appended to any OLS model in order to make the exposure coefficient technically unbiased across realizations of the random effect. 

\section{Illustration of the difference between a linear coefficient and an average of 1-unit shifts}\label{sec:linear}

\begin{figure}
	\centering
	\includegraphics[scale=0.5]{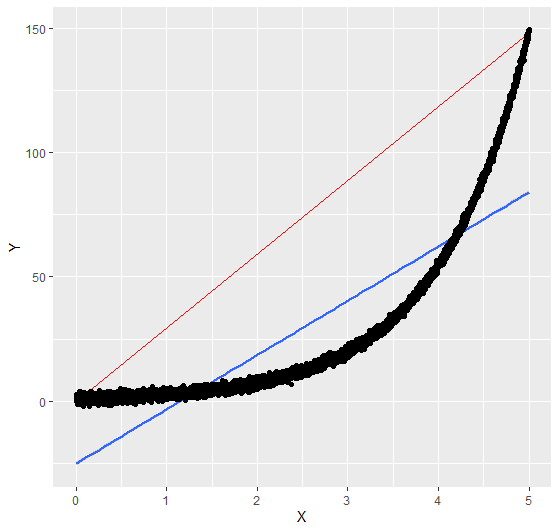}
	\caption{The red line has slope equal to the average incremental effect of intervention on exposure; the blue line is the OLS fit.}
	\label{fig:nonlinear}
\end{figure}

See Figure \ref{fig:nonlinear}).
We generate data from %
$X \sim \unif(0, 5)$ and 
$Y \sim N(\exp(X), 1)$. 
In this case, the average effect of an incremental intervention (that is, the average slope of the exposure-response curve) on $X$ may be calculated as
\begin{gather*}
	E[ \frac{dE[Y]}{dX} ] = E[\exp X] = \int_0^5 e^x dx / 5  \approx 29.5
\end{gather*}
But linear regression will approximately minimize $\int_0^5 (e^x - (\beta_0 + \beta_1 x))^2 dx$, resulting in an effect estimate of $21.7$. %

\section{Simulation results}\label{app_sim_res}

\subsection{Details of the simulation studies}\label{sec:simdetails}
The locations are generated uniformly at random on the $[-1,1]\times[-1,1]$ square, and new locations are drawn in every replicate. In the small-sample simulations, 1,000 observations are generated for 500 replicates. In the large-sample simulations, 10,000 observations are simulated for 250 replicates. By design, the true effect in all simulations except the non-linear effect scenario is 1. In the non-linear effect scenario, the true effect was assessed by simulation to be 2.431. %

The methods we compared are described in Table \ref{tab:methods}.

\begin{footnotesize}
	\begin{table}[!t]
		\centering
		\begin{tabular}{p{0.6\linewidth} | p{0.35\linewidth}}
			\hline
			Method & Interval estimation procedure\\
			\hline
			OLS / RSR \citep{hodges} (unadajusted estimator) & Bootstrap. \\ \hline
			spline: PLM with spatial effect modeled by thin plate regression spline \citep{wood}. Implemented by \texttt{mgcv} \citep{mgcv}. & Analytic confidence intervals.\\ \hline
			gp: PLM with spatial effect modeled by two-dimensional nearest-neighbors Gaussian process (NNGP) . Implemented by \texttt{BRISC} package \citep{BRISC}. & BRISC bootstrap \citep{saha}.\\ \hline
			gSEM: geo-additive structural equation model \citep{thaden}. Splines fit by \texttt{mgcv}. & Bootstrap.\\ \hline
			spatial+: spatial+ \citep{dupont}. Splines fit by \texttt{mgcv}. & Bootstrap. \\ \hline
			svc\_mle: Spatially-varying coefficient model \citep{dambon}. Implemented by \texttt{varycoef} package \citep{varycoef}. & Omitted (since resampled points cannot be accommodated by the GP covariance function). \\ \hline
			spline\_interaction: Spline with spatial interaction of exposure and space (hence, a spatially-varying coefficient model). Implemented by \texttt{mgcv}. & Bootstrap. \\ \hline
			gp3: Three-dimensional Gaussian process over location and exposure (flexible model). Implemented by \texttt{RobustGaSP} \citep{RobustGaSP}. & Omitted (since resampled points cannot be accommodated by the GP covariance function). \\ \hline
			spline3: Three-dimensional spline over location and exposure (flexible model). Implemented by \texttt{mgcv}. & Omitted due to computational constraints.\\ \hline
			rf: Random forest over location and exposure (flexible model) \citep{breiman}. Implemented by \texttt{randomForest} package \citep{randomForest}. & Bootstrap.\\ \hline
			BART: Bayesian additive regression tree (BART) over location and exposure (flexible model) \citep{chipman}. Implemented by \texttt{dbarts} \citep{dbarts} package. &  Bootstrap.\\ \hline
			DML\_gp3: gp3 model augmented with spline model over space for exposure (flexible model, DML). & Omitted (since resampled points cannot be accommodated by the GP covariance function).\\ \hline
			DML\_spline: spline3 model augmented with spline model over space for exposure (flexible model, DML). & Omitted due to computational constraints. \\ \hline
			DML\_BART: BART model augmented with spline model over space for exposure (flexible model, DML). & Bootstrap. \\
			\hline
		\end{tabular}
		\caption{Methods considered for the simulation studies.}
	\end{table}
	\label{tab:methods}
\end{footnotesize}

Confidence intervals are for 95\% nominal coverage. Bootstrap confidence intervals use 120 resamples to obtain a normal approximation to the sampling distribution, as recommended in \cite{efron}; that is, the standard deviation of the bootstrap sampling distribution is multiplied by 1.96 to form the half-width of the confidence intervals centered at the empirical estimate. 

For linear models, we use the linear coefficient estimate to estimate the effect of a $+1$ shift intervention. For other models, other than the DML methods, we use the model to predict potential outcomes under the shift intervention and then take the average difference with the model's fitted observed values. 

We use package-default settings in general, but we adjust the parameter $k$ in the \texttt{mgcv} package which controls the maximum number of degrees of freedom in the spline fits so that they can adequately approximate complicated functions. For splines over two dimensions, we use $k=200$. For three-dimensional splines, we use $k=1000$ in the large sample and $k=500$ in the small sample.

To estimate the effect of the shift intervention by DML, we implement the procedure described in the Supplementary Material Section \ref{app_hr}. While the procedure for the regression model varies, for each method we fit the exposure model with a thin plate regression spline as $X_i = g(S_i) + \epsilon_i$ where the errors are i.i.d. The i.i.d. assumption simplifies the estimation of the function $\lambda$ (by reducing the conditional density estimation to an unconditional density estimation), but the assumption is violated in the heterogeneous effect simulations (in which case errors are independent but not identically distributed). 

\subsection{Linear confounding}

\begin{gather*}
	S^a, S^b \sim_{i.i.d.} \unif(-1, 1)\\
	U =  S^a + S^b\\
	X \sim N(U, 5)\\
	Y \sim N(3U+X, 1)
\end{gather*}

\clearpage
\begin{table}[ht]
	\begin{adjustwidth}{-3cm}{-3cm}
		\centering
		\begin{tabular}{rlrrrl}
			\hline
			n & method & bias & sd & mse & coverage \\ 
			\hline
			$1.000 \times 10^{3}$ & RSR & $7.733 \times 10^{-2}$ & $1.603 \times 10^{-2}$ & $6.237 \times 10^{-3}$ & 0\% \\ 
			$1.000 \times 10^{3}$ & spline & $-9.958 \times 10^{-5}$ & $6.297 \times 10^{-3}$ & $\mathbf{3.958 \times 10^{-5}}$ & 96\% \\ 
			$1.000 \times 10^{3}$ & gp & $7.932 \times 10^{-4}$ & $6.509 \times 10^{-3}$ & $4.292 \times 10^{-5}$ & 95\% \\ 
			$1.000 \times 10^{3}$ & gSEM & $1.362 \times 10^{-3}$ & $7.017 \times 10^{-3}$ & $5.099 \times 10^{-5}$ & 99\% \\ 
			$1.000 \times 10^{3}$ & spatial+ & $5.348 \times 10^{-4}$ & $6.477 \times 10^{-3}$ & $4.216 \times 10^{-5}$ & 98\% \\ 
			$1.000 \times 10^{3}$ & svc\_mle & $3.294 \times 10^{-4}$ & $6.434 \times 10^{-3}$ & $4.143 \times 10^{-5}$ &  \\ 
			$1.000 \times 10^{3}$ & spline\_interaction & $-9.566 \times 10^{-5}$ & $6.338 \times 10^{-3}$ & $4.010 \times 10^{-5}$ & 99\% \\ 
			$1.000 \times 10^{3}$ & gp3 & $-5.282 \times 10^{-4}$ & $6.478 \times 10^{-3}$ & $4.216 \times 10^{-5}$ &  \\ 
			$1.000 \times 10^{3}$ & spline3 & $\mathbf{1.538 \times 10^{-5}}$ & $6.539 \times 10^{-3}$ & $4.268 \times 10^{-5}$ &  \\ 
			$1.000 \times 10^{3}$ & rf & $-1.886 \times 10^{-1}$ & $1.370 \times 10^{-2}$ & $3.575 \times 10^{-2}$ & 0\% \\ 
			$1.000 \times 10^{3}$ & BART & $-2.754 \times 10^{-3}$ & $1.340 \times 10^{-2}$ & $1.868 \times 10^{-4}$ & 100\% \\ 
			$1.000 \times 10^{3}$ & DML\_gp3 & $-2.531 \times 10^{-4}$ & $6.690 \times 10^{-3}$ & $4.473 \times 10^{-5}$ &  \\ 
			$1.000 \times 10^{3}$ & DML\_spline & $1.424 \times 10^{-4}$ & $6.749 \times 10^{-3}$ & $4.548 \times 10^{-5}$ &  \\ 
			$1.000 \times 10^{3}$ & DML\_BART & $-1.811 \times 10^{-3}$ & $1.350 \times 10^{-2}$ & $1.850 \times 10^{-4}$ & 100\% \\ \hline 
			$1.000 \times 10^{4}$ & RSR & $7.715 \times 10^{-2}$ & $5.237 \times 10^{-3}$ & $5.980 \times 10^{-3}$ & 0\% \\ 
			$1.000 \times 10^{4}$ & spline & $-2.203 \times 10^{-4}$ & $1.914 \times 10^{-3}$ & $3.696 \times 10^{-6}$ & 96\% \\ 
			$1.000 \times 10^{4}$ & gp & $-1.842 \times 10^{-4}$ & $1.978 \times 10^{-3}$ & $3.932 \times 10^{-6}$ & 96\% \\ 
			$1.000 \times 10^{4}$ & gSEM & $\mathbf{3.427 \times 10^{-6}}$ & $1.934 \times 10^{-3}$ & $3.724 \times 10^{-6}$ & 98\% \\ 
			$1.000 \times 10^{4}$ & spatial+ & $-1.298 \times 10^{-4}$ & $1.912 \times 10^{-3}$ & $\mathbf{3.657 \times 10^{-6}}$ & 97\% \\ 
			$1.000 \times 10^{4}$ & spline\_interaction & $-2.171 \times 10^{-4}$ & $1.911 \times 10^{-3}$ & $3.684 \times 10^{-6}$ & 97\% \\ 
			$1.000 \times 10^{4}$ & spline3 & $-1.965 \times 10^{-4}$ & $1.935 \times 10^{-3}$ & $3.768 \times 10^{-6}$ &  \\ 
			$1.000 \times 10^{4}$ & rf & $-9.426 \times 10^{-2}$ & $4.410 \times 10^{-3}$ & $8.904 \times 10^{-3}$ & 0\% \\ 
			$1.000 \times 10^{4}$ & BART & $-8.981 \times 10^{-4}$ & $2.987 \times 10^{-3}$ & $9.694 \times 10^{-6}$ & 98\% \\ 
			$1.000 \times 10^{4}$ & DML\_spline & $-1.965 \times 10^{-4}$ & $1.991 \times 10^{-3}$ & $3.987 \times 10^{-6}$ &  \\ 
			$1.000 \times 10^{4}$ & DML\_BART & $-3.668 \times 10^{-4}$ & $2.992 \times 10^{-3}$ & $9.053 \times 10^{-6}$ & 98\% \\ 
			\hline
		\end{tabular}
		\caption{Simulation results in the linear confounding scenario}
		\label{tab:sim_linear}
	\end{adjustwidth}
\end{table}

\subsection{Simple effect}
\begin{gather*}
	S^a, S^b \sim_{i.i.d.} \unif(-1, 1)\\
	U = \sin(2\pi S^a*S^b) + S^a + S^b\\
	X \sim N(U^3, 5)\\
	Y \sim N(3U+X, 1)
\end{gather*}
\clearpage
\begin{table}[ht]
	\begin{adjustwidth}{-3cm}{-3cm}
		\centering
		\begin{tabular}{rlrrrr}
			\hline
			n & method & bias & sd & mse & coverage \\ 
			\hline
			$1.000 \times 10^{3}$ & RSR & $2.589 \times 10^{-1}$ & $1.528 \times 10^{-2}$ & $6.727 \times 10^{-2}$ & 0\% \\ 
			$1.000 \times 10^{3}$ & spline & $6.863 \times 10^{-3}$ & $6.502 \times 10^{-3}$ & $8.930 \times 10^{-5}$ & 81\% \\ 
			$1.000 \times 10^{3}$ & gp & $7.727 \times 10^{-3}$ & $6.770 \times 10^{-3}$ & $1.054 \times 10^{-4}$ & 76\% \\ 
			$1.000 \times 10^{3}$ & gSEM & $7.305 \times 10^{-2}$ & $7.584 \times 10^{-3}$ & $5.394 \times 10^{-3}$ & 0\% \\ 
			$1.000 \times 10^{3}$ & spatial+ & $1.084 \times 10^{-2}$ & $6.826 \times 10^{-3}$ & $1.641 \times 10^{-4}$ & 73\% \\ 
			$1.000 \times 10^{3}$ & svc\_mle & $7.387 \times 10^{-3}$ & $6.855 \times 10^{-3}$ & $1.015 \times 10^{-4}$ &  \\ 
			$1.000 \times 10^{3}$ & spline\_interaction & $6.801 \times 10^{-3}$ & $6.568 \times 10^{-3}$ & $8.930 \times 10^{-5}$ & 96\% \\ 
			$1.000 \times 10^{3}$ & gp3 & $2.682 \times 10^{-3}$ & $6.939 \times 10^{-3}$ & $5.524 \times 10^{-5}$ &  \\ 
			$1.000 \times 10^{3}$ & spline3 & $6.308 \times 10^{-2}$ & $4.068 \times 10^{-2}$ & $5.631 \times 10^{-3}$ &  \\ 
			$1.000 \times 10^{3}$ & rf & $-1.458 \times 10^{-1}$ & $1.595 \times 10^{-2}$ & $2.152 \times 10^{-2}$ & 0\% \\ 
			$1.000 \times 10^{3}$ & BART & $7.672 \times 10^{-3}$ & $1.680 \times 10^{-2}$ & $3.405 \times 10^{-4}$ & 100\% \\ 
			$1.000 \times 10^{3}$ & DML\_gp3 & $\mathbf{1.396 \times 10^{-3}}$ & $7.065 \times 10^{-3}$ & $\mathbf{5.177 \times 10^{-5}}$ &  \\ 
			$1.000 \times 10^{3}$ & DML\_spline & $2.998 \times 10^{-2}$ & $3.476 \times 10^{-2}$ & $2.104 \times 10^{-3}$ &  \\ 
			$1.000 \times 10^{3}$ & DML\_BART & $3.432 \times 10^{-3}$ & $1.685 \times 10^{-2}$ & $2.950 \times 10^{-4}$ & 100\% \\ \hline 
			$1.000 \times 10^{4}$ & RSR & $2.580 \times 10^{-1}$ & $4.749 \times 10^{-3}$ & $6.661 \times 10^{-2}$ & 0\% \\ 
			$1.000 \times 10^{4}$ & spline & $1.630 \times 10^{-3}$ & $1.947 \times 10^{-3}$ & $6.433 \times 10^{-6}$ & 86\% \\ 
			$1.000 \times 10^{4}$ & gp & $2.159 \times 10^{-3}$ & $1.990 \times 10^{-3}$ & $8.605 \times 10^{-6}$ & 81\% \\ 
			$1.000 \times 10^{4}$ & gSEM & $1.564 \times 10^{-2}$ & $2.089 \times 10^{-3}$ & $2.488 \times 10^{-4}$ & 0\% \\ 
			$1.000 \times 10^{4}$ & spatial+ & $5.063 \times 10^{-4}$ & $1.958 \times 10^{-3}$ & $\mathbf{4.075 \times 10^{-6}}$ & 96\% \\ 
			$1.000 \times 10^{4}$ & spline\_interaction & $1.614 \times 10^{-3}$ & $1.945 \times 10^{-3}$ & $6.372 \times 10^{-6}$ & 88\% \\ 
			$1.000 \times 10^{4}$ & spline3 & $2.244 \times 10^{-2}$ & $6.780 \times 10^{-3}$ & $5.495 \times 10^{-4}$ &  \\ 
			$1.000 \times 10^{4}$ & rf & $-7.677 \times 10^{-2}$ & $4.497 \times 10^{-3}$ & $5.914 \times 10^{-3}$ & 0\% \\ 
			$1.000 \times 10^{4}$ & BART & $2.642 \times 10^{-3}$ & $3.618 \times 10^{-3}$ & $2.002 \times 10^{-5}$ & 97\% \\ 
			$1.000 \times 10^{4}$ & DML\_spline & $6.870 \times 10^{-3}$ & $7.080 \times 10^{-3}$ & $9.712 \times 10^{-5}$ &  \\ 
			$1.000 \times 10^{4}$ & DML\_BART & $\mathbf{4.626 \times 10^{-4}}$ & $3.586 \times 10^{-3}$ & $1.302 \times 10^{-5}$ & 100\% \\ 
			\hline
		\end{tabular}
		\caption{Simulation results in the simple effect scenario}
		\label{tab:sim_simple}
	\end{adjustwidth}
\end{table}

\subsection{Spatially structured heterogeneous effect}
\begin{gather*}
	S^a, S^b \sim_{i.i.d.} \unif(-1, 1)\\
	U = \sin(2\pi S^a*S^b) + S^a + S^b\\
	X \sim N(U^3, 5*\exp(U/2))\\
	Y \sim N(3U+(1+U)*X, 1)
\end{gather*}
\clearpage

\begin{table}[ht]
	\begin{adjustwidth}{-3cm}{-3cm}
		\centering
		\begin{tabular}{rlrrrl}
			\hline
			n & method & bias & sd & mse & coverage \\ 
			\hline
			$1.000 \times 10^{3}$ & RSR & $8.670 \times 10^{-1}$ & $1.173 \times 10^{-1}$ & $7.655 \times 10^{-1}$ & 0\% \\ 
			$1.000 \times 10^{3}$ & spline & $1.070 \times 10^{0}$ & $7.224 \times 10^{-2}$ & $1.150 \times 10^{0}$ & 0\% \\ 
			$1.000 \times 10^{3}$ & gp & $1.063 \times 10^{0}$ & $7.418 \times 10^{-2}$ & $1.135 \times 10^{0}$ & 0\% \\ 
			$1.000 \times 10^{3}$ & gSEM & $1.123 \times 10^{0}$ & $7.587 \times 10^{-2}$ & $1.268 \times 10^{0}$ & 0\% \\ 
			$1.000 \times 10^{3}$ & spatial+ & $1.090 \times 10^{0}$ & $7.215 \times 10^{-2}$ & $1.193 \times 10^{0}$ & 0\% \\ 
			$1.000 \times 10^{3}$ & svc\_mle & $2.019 \times 10^{-2}$ & $3.384 \times 10^{-2}$ & $1.551 \times 10^{-3}$ &  \\ 
			$1.000 \times 10^{3}$ & spline\_interaction & $1.390 \times 10^{-2}$ & $3.400 \times 10^{-2}$ & $1.347 \times 10^{-3}$ & 94\% \\ 
			$1.000 \times 10^{3}$ & gp3 & $3.863 \times 10^{-3}$ & $3.541 \times 10^{-2}$ & $1.266 \times 10^{-3}$ &  \\ 
			$1.000 \times 10^{3}$ & spline3 & $1.545 \times 10^{-1}$ & $1.074 \times 10^{-1}$ & $3.539 \times 10^{-2}$ &  \\ 
			$1.000 \times 10^{3}$ & rf & $-7.837 \times 10^{-2}$ & $4.106 \times 10^{-2}$ & $7.824 \times 10^{-3}$ & 67\% \\ 
			$1.000 \times 10^{3}$ & BART & $1.024 \times 10^{-1}$ & $7.197 \times 10^{-2}$ & $1.566 \times 10^{-2}$ & 90\% \\ 
			$1.000 \times 10^{3}$ & DML\_gp3 & $\mathbf{3.363 \times 10^{-3}}$ & $3.541 \times 10^{-2}$ & $\mathbf{1.262 \times 10^{-3}}$ &  \\ 
			$1.000 \times 10^{3}$ & DML\_spline & $1.439 \times 10^{-1}$ & $1.073 \times 10^{-1}$ & $3.220 \times 10^{-2}$ &  \\ 
			$1.000 \times 10^{3}$ & DML\_BART & $9.417 \times 10^{-2}$ & $7.194 \times 10^{-2}$ & $1.403 \times 10^{-2}$ & 92\% \\ \hline
			$1.000 \times 10^{4}$ & RSR & $8.687 \times 10^{-1}$ & $4.060 \times 10^{-2}$ & $7.562 \times 10^{-1}$ & 0\% \\ 
			$1.000 \times 10^{4}$ & spline & $1.088 \times 10^{0}$ & $2.393 \times 10^{-2}$ & $1.183 \times 10^{0}$ & 0\% \\ 
			$1.000 \times 10^{4}$ & gp & $1.085 \times 10^{0}$ & $2.416 \times 10^{-2}$ & $1.178 \times 10^{0}$ & 0\% \\ 
			$1.000 \times 10^{4}$ & gSEM & $1.094 \times 10^{0}$ & $2.423 \times 10^{-2}$ & $1.197 \times 10^{0}$ & 0\% \\ 
			$1.000 \times 10^{4}$ & spatial+ & $1.090 \times 10^{0}$ & $2.389 \times 10^{-2}$ & $1.189 \times 10^{0}$ & 0\% \\ 
			$1.000 \times 10^{4}$ & spline\_interaction & $\mathbf{2.216 \times 10^{-3}}$ & $1.080 \times 10^{-2}$ & $\mathbf{1.210 \times 10^{-4}}$ & 94\% \\ 
			$1.000 \times 10^{4}$ & spline3 & $8.317 \times 10^{-2}$ & $2.091 \times 10^{-2}$ & $7.353 \times 10^{-3}$ &  \\ 
			$1.000 \times 10^{4}$ & rf & $-4.038 \times 10^{-2}$ & $1.200 \times 10^{-2}$ & $1.774 \times 10^{-3}$ & 16\% \\ 
			$1.000 \times 10^{4}$ & BART & $3.066 \times 10^{-2}$ & $1.514 \times 10^{-2}$ & $1.168 \times 10^{-3}$ & 67\% \\ 
			$1.000 \times 10^{4}$ & DML\_spline & $6.955 \times 10^{-2}$ & $2.175 \times 10^{-2}$ & $5.308 \times 10^{-3}$ &  \\ 
			$1.000 \times 10^{4}$ & DML\_BART & $2.832 \times 10^{-2}$ & $1.504 \times 10^{-2}$ & $1.027 \times 10^{-3}$ & 74\% \\ 
			\hline
		\end{tabular}    \caption{Spatially-structured heterogeneous effect simulation results}
		\label{tab:sim_structhet}
	\end{adjustwidth} 
\end{table}    

\subsection{Non-linear effect}

\begin{gather*}
	S^a, S^b \sim_{i.i.d.} \unif(-1, 1)\\
	U = \sin(2\pi S^a*S^b) + S^a + S^b\\
	X \sim N(U^3, 5)\\
	Y \sim N(3U+X+X^2, 1)
\end{gather*}
\clearpage
\begin{table}[ht]
	\begin{adjustwidth}{-3cm}{-3cm}
		\centering
		\begin{tabular}{rlrrrr}
			\hline
			n & method & bias & sd & mse & coverage \\ 
			\hline
			$1.000 \times 10^{3}$ & RSR & $-2.791 \times 10^{0}$ & $1.065 \times 10^{0}$ & $8.921 \times 10^{0}$ & 0\% \\ 
			$1.000 \times 10^{3}$ & spline & $-1.157 \times 10^{0}$ & $6.826 \times 10^{-1}$ & $1.804 \times 10^{0}$ & 25\% \\ 
			$1.000 \times 10^{3}$ & gp & $-1.322 \times 10^{0}$ & $6.891 \times 10^{-1}$ & $2.221 \times 10^{0}$ & 17\% \\ 
			$1.000 \times 10^{3}$ & gSEM & $-1.254 \times 10^{0}$ & $7.258 \times 10^{-1}$ & $2.098 \times 10^{0}$ & 55\% \\ 
			$1.000 \times 10^{3}$ & spatial+ & $-1.087 \times 10^{0}$ & $6.927 \times 10^{-1}$ & $1.661 \times 10^{0}$ & 63\% \\ 
			$1.000 \times 10^{3}$ & svc\_mle & $-1.145 \times 10^{0}$ & $2.002 \times 10^{-1}$ & $1.350 \times 10^{0}$ &  \\ 
			$1.000 \times 10^{3}$ & spline\_interaction & $-9.852 \times 10^{-1}$ & $4.471 \times 10^{-1}$ & $1.170 \times 10^{0}$ & 38\% \\ 
			$1.000 \times 10^{3}$ & gp3 & $3.652 \times 10^{-2}$ & $3.773 \times 10^{-1}$ & $1.434 \times 10^{-1}$ &  \\ 
			$1.000 \times 10^{3}$ & spline3 & $8.767 \times 10^{-2}$ & $3.791 \times 10^{-1}$ & $1.511 \times 10^{-1}$ &  \\ 
			$1.000 \times 10^{3}$ & rf & $-3.678 \times 10^{-1}$ & $3.409 \times 10^{-1}$ & $2.513 \times 10^{-1}$ & 75\% \\ 
			$1.000 \times 10^{3}$ & BART & $2.818 \times 10^{-2}$ & $3.902 \times 10^{-1}$ & $1.527 \times 10^{-1}$ & 98\% \\ 
			$1.000 \times 10^{3}$ & DML\_gp3 & $2.911 \times 10^{-2}$ & $3.771 \times 10^{-1}$ & $\mathbf{1.428 \times 10^{-1}}$ &  \\ 
			$1.000 \times 10^{3}$ & DML\_spline & $5.456 \times 10^{-2}$ & $3.765 \times 10^{-1}$ & $1.444 \times 10^{-1}$ &  \\ 
			$1.000 \times 10^{3}$ & DML\_BART & $\mathbf{4.759 \times 10^{-3}}$ & $3.901 \times 10^{-1}$ & $1.519 \times 10^{-1}$ & 98\% \\ \hline
			$1.000 \times 10^{4}$ & RSR & $-2.853 \times 10^{0}$ & $3.774 \times 10^{-1}$ & $8.283 \times 10^{0}$ & 0\% \\ 
			$1.000 \times 10^{4}$ & spline & $-1.061 \times 10^{0}$ & $2.014 \times 10^{-1}$ & $1.166 \times 10^{0}$ & 0\% \\ 
			$1.000 \times 10^{4}$ & gp & $-1.084 \times 10^{0}$ & $2.010 \times 10^{-1}$ & $1.215 \times 10^{0}$ & 0\% \\ 
			$1.000 \times 10^{4}$ & gSEM & $-1.146 \times 10^{0}$ & $2.053 \times 10^{-1}$ & $1.356 \times 10^{0}$ & 0\% \\ 
			$1.000 \times 10^{4}$ & spatial+ & $-1.042 \times 10^{0}$ & $2.014 \times 10^{-1}$ & $1.126 \times 10^{0}$ & 0\% \\ 
			$1.000 \times 10^{4}$ & spline\_interaction & $-9.893 \times 10^{-1}$ & $1.594 \times 10^{-1}$ & $1.004 \times 10^{0}$ & 0\% \\ 
			$1.000 \times 10^{4}$ & spline3 & $2.979 \times 10^{-2}$ & $1.229 \times 10^{-1}$ & $1.594 \times 10^{-2}$ &  \\ 
			$1.000 \times 10^{4}$ & rf & $-1.456 \times 10^{-1}$ & $1.143 \times 10^{-1}$ & $3.421 \times 10^{-2}$ & 75\% \\ 
			$1.000 \times 10^{4}$ & BART & $1.308 \times 10^{-2}$ & $1.253 \times 10^{-1}$ & $1.581 \times 10^{-2}$ & 96\% \\ 
			$1.000 \times 10^{4}$ & DML\_spline & $1.421 \times 10^{-2}$ & $1.229 \times 10^{-1}$ & $\mathbf{1.524 \times 10^{-2}}$ &  \\ 
			$1.000 \times 10^{4}$ & DML\_BART & $\mathbf{5.706 \times 10^{-3}}$ & $1.254 \times 10^{-1}$ & $1.569 \times 10^{-2}$ & 96\% \\ \hline
		\end{tabular}
		\caption{Non-linear effect simulation results}
		\label{tab:sim_nonlin}
	\end{adjustwidth}
\end{table}

\subsection{Random heterogeneous effect}
\begin{gather*}
	S^a, S^b \sim_{i.i.d.} \unif(-1, 1)\\
	U = \sin(2\pi S^a*S^b) + S^a + S^b\\
	X \sim N(U^3, 5*\exp(U/2))\\
	d \sim N(0,1)\\
	Y \sim N(U+(1+d)*X, 1)
\end{gather*}
\clearpage

\begin{table}[ht]
	\begin{adjustwidth}{-3cm}{-3cm}
		\centering
		\begin{tabular}{rlrrrr}
			\hline
			n & method & bias & sd & mse & coverage \\ 
			\hline
			$1.000 \times 10^{3}$ & RSR & $1.731 \times 10^{-1}$ & $8.450 \times 10^{-2}$ & $3.707 \times 10^{-2}$ & 0\% \\ 
			$1.000 \times 10^{3}$ & spline & $2.433 \times 10^{-2}$ & $8.790 \times 10^{-2}$ & $8.302 \times 10^{-3}$ & 51\% \\ 
			$1.000 \times 10^{3}$ & gp & $3.795 \times 10^{-2}$ & $8.729 \times 10^{-2}$ & $9.044 \times 10^{-3}$ & 51\% \\ 
			$1.000 \times 10^{3}$ & gSEM & $5.425 \times 10^{-2}$ & $9.213 \times 10^{-2}$ & $1.141 \times 10^{-2}$ & 90\% \\ 
			$1.000 \times 10^{3}$ & spatial+ & $2.323 \times 10^{-2}$ & $9.081 \times 10^{-2}$ & $8.770 \times 10^{-3}$ & 93\% \\ 
			$1.000 \times 10^{3}$ & svc\_mle & $2.079 \times 10^{-2}$ & $3.639 \times 10^{-2}$ & $\mathbf{1.754 \times 10^{-3}}$ &  \\ 
			$1.000 \times 10^{3}$ & spline\_interaction & $4.651 \times 10^{-2}$ & $5.233 \times 10^{-2}$ & $4.896 \times 10^{-3}$ & 94\% \\ 
			$1.000 \times 10^{3}$ & gp3 & $1.817 \times 10^{-2}$ & $1.390 \times 10^{-1}$ & $1.962 \times 10^{-2}$ &  \\ 
			$1.000 \times 10^{3}$ & spline3 & $4.436 \times 10^{-2}$ & $2.268 \times 10^{-1}$ & $5.331 \times 10^{-2}$ &  \\ 
			$1.000 \times 10^{3}$ & rf & $-1.103 \times 10^{-1}$ & $5.951 \times 10^{-2}$ & $1.571 \times 10^{-2}$ & 72\% \\ 
			$1.000 \times 10^{3}$ & BART & $1.216 \times 10^{-2}$ & $1.014 \times 10^{-1}$ & $1.040 \times 10^{-2}$ & 100\% \\ 
			$1.000 \times 10^{3}$ & DML\_gp3 & $-1.420 \times 10^{-2}$ & $1.363 \times 10^{-1}$ & $1.874 \times 10^{-2}$ &  \\ 
			$1.000 \times 10^{3}$ & DML\_spline & $2.872 \times 10^{-2}$ & $2.301 \times 10^{-1}$ & $5.369 \times 10^{-2}$ &  \\ 
			$1.000 \times 10^{3}$ & DML\_BART & $ \mathbf{6.093 \times 10^{-3}}$ & $1.023 \times 10^{-1}$ & $1.048 \times 10^{-2}$ & 100\% \\ \hline
			$1.000 \times 10^{4}$ & RSR & $1.760 \times 10^{-1}$ & $2.928 \times 10^{-2}$ & $3.184 \times 10^{-2}$ & 0\% \\ 
			$1.000 \times 10^{4}$ & spline & $1.043 \times 10^{-2}$ & $2.919 \times 10^{-2}$ & $9.573 \times 10^{-4}$ & 52\% \\ 
			$1.000 \times 10^{4}$ & gp & $2.034 \times 10^{-2}$ & $2.946 \times 10^{-2}$ & $1.278 \times 10^{-3}$ & 47\% \\ 
			$1.000 \times 10^{4}$ & gSEM & $1.407 \times 10^{-2}$ & $2.939 \times 10^{-2}$ & $1.058 \times 10^{-3}$ & 90\% \\ 
			$1.000 \times 10^{4}$ & spatial+ & $5.516 \times 10^{-3}$ & $2.923 \times 10^{-2}$ & $8.816 \times 10^{-4}$ & 93\% \\ 
			$1.000 \times 10^{4}$ & spline\_interaction & $2.210 \times 10^{-2}$ & $1.800 \times 10^{-2}$ & $\mathbf{8.110 \times 10^{-4}}$ & 80\% \\ 
			$1.000 \times 10^{4}$ & spline3 & $4.608 \times 10^{-2}$ & $2.875 \times 10^{-2}$ & $2.947 \times 10^{-3}$ &  \\ 
			$1.000 \times 10^{4}$ & rf & $-5.237 \times 10^{-2}$ & $2.216 \times 10^{-2}$ & $3.231 \times 10^{-3}$ & 62\% \\ 
			$1.000 \times 10^{4}$ & BART & $1.036 \times 10^{-2}$ & $3.330 \times 10^{-2}$ & $1.212 \times 10^{-3}$ & 100\% \\ 
			$1.000 \times 10^{4}$ & DML\_spline & $1.740 \times 10^{-2}$ & $3.327 \times 10^{-2}$ & $1.405 \times 10^{-3}$ &  \\ 
			$1.000 \times 10^{4}$ & DML\_BART & $\mathbf{1.913 \times 10^{-3}}$ & $3.456 \times 10^{-2}$ & $1.193 \times 10^{-3}$ & 100\% \\ 
			\hline
		\end{tabular}
		\caption{Random heterogeneous effect simulation results}
		\label{tab:sim_randhet}
	\end{adjustwidth}
\end{table}

\subsection{Noisy confounding}\label{sec:noisy}
\begin{gather*}
	S^a, S^b \sim_{i.i.d.} \unif(-1, 1)\\
	U \sim N( \sin(2\pi S^a*S^b) + S^a + S^b, 1)\\
	X \sim N(U^3, 5)\\
	Y \sim N(5U+X, 1)
\end{gather*}

\clearpage

\begin{table}[ht]
	\begin{adjustwidth}{-3cm}{-3cm}
		\centering
		\begin{tabular}{rlrrrr}
			\hline
			n & method & bias & sd & mse & coverage \\ 
			\hline
			$1.000 \times 10^{3}$ & RSR & $4.240 \times 10^{-1}$ & $3.038 \times 10^{-2}$ & $1.807 \times 10^{-1}$ & 0\% \\ 
			$1.000 \times 10^{3}$ & spline & $2.979 \times 10^{-1}$ & $2.432 \times 10^{-2}$ & $8.936 \times 10^{-2}$ & 0\% \\ 
			$1.000 \times 10^{3}$ & gp & $3.004 \times 10^{-1}$ & $2.442 \times 10^{-2}$ & $9.086 \times 10^{-2}$ & 0\% \\ 
			$1.000 \times 10^{3}$ & gSEM & $3.521 \times 10^{-1}$ & $2.521 \times 10^{-2}$ & $1.246 \times 10^{-1}$ & 0\% \\ 
			$1.000 \times 10^{3}$ & spatial+ & $3.055 \times 10^{-1}$ & $2.443 \times 10^{-2}$ & $9.391 \times 10^{-2}$ & 0\% \\ 
			$1.000 \times 10^{3}$ & svc\_mle & $3.397 \times 10^{-1}$ & $2.318 \times 10^{-2}$ & $1.160 \times 10^{-1}$ &  \\ 
			$1.000 \times 10^{3}$ & spline\_interaction & $3.436 \times 10^{-1}$ & $1.890 \times 10^{-2}$ & $1.184 \times 10^{-1}$ & 0\% \\ 
			$1.000 \times 10^{3}$ & gp3 & $3.753 \times 10^{-1}$ & $2.096 \times 10^{-2}$ & $1.413 \times 10^{-1}$ &  \\ 
			$1.000 \times 10^{3}$ & spline3 & $4.230 \times 10^{-1}$ & $2.279 \times 10^{-2}$ & $1.795 \times 10^{-1}$ &  \\ 
			$1.000 \times 10^{3}$ & rf & $\mathbf{1.072 \times 10^{-1}}$ & $3.980 \times 10^{-2}$ & $\mathbf{1.308 \times 10^{-2}}$ & 49\% \\ 
			$1.000 \times 10^{3}$ & BART & $3.278 \times 10^{-1}$ & $5.504 \times 10^{-2}$ & $1.105 \times 10^{-1}$ & 0\% \\ 
			$1.000 \times 10^{3}$ & DML\_gp3 & $3.572 \times 10^{-1}$ & $2.244 \times 10^{-2}$ & $1.281 \times 10^{-1}$ &  \\ 
			$1.000 \times 10^{3}$ & DML\_spline & $3.324 \times 10^{-1}$ & $2.567 \times 10^{-2}$ & $1.111 \times 10^{-1}$ &  \\ 
			$1.000 \times 10^{3}$ & DML\_BART & $3.204 \times 10^{-1}$ & $5.549 \times 10^{-2}$ & $1.057 \times 10^{-1}$ & 0\% \\ \hline
			$1.000 \times 10^{4}$ & RSR & $4.197 \times 10^{-1}$ & $1.015 \times 10^{-2}$ & $1.763 \times 10^{-1}$ & 0\% \\ 
			$1.000 \times 10^{4}$ & spline & $2.879 \times 10^{-1}$ & $7.850 \times 10^{-3}$ & $8.295 \times 10^{-2}$ & 0\% \\ 
			$1.000 \times 10^{4}$ & gp & $2.880 \times 10^{-1}$ & $7.820 \times 10^{-3}$ & $8.299 \times 10^{-2}$ & 0\% \\ 
			$1.000 \times 10^{4}$ & gSEM & $3.007 \times 10^{-1}$ & $8.043 \times 10^{-3}$ & $9.051 \times 10^{-2}$ & 0\% \\ 
			$1.000 \times 10^{4}$ & spatial+ & $2.879 \times 10^{-1}$ & $7.798 \times 10^{-3}$ & $8.296 \times 10^{-2}$ & 0\% \\ 
			$1.000 \times 10^{4}$ & spline\_interaction & $3.443 \times 10^{-1}$ & $5.832 \times 10^{-3}$ & $1.186 \times 10^{-1}$ & 0\% \\ 
			$1.000 \times 10^{4}$ & spline3 & $3.678 \times 10^{-1}$ & $1.214 \times 10^{-2}$ & $1.354 \times 10^{-1}$ &  \\ 
			$1.000 \times 10^{4}$ & rf & $\mathbf{2.182 \times 10^{-1}}$ & $1.239 \times 10^{-2}$ & $\mathbf{4.775 \times 10^{-2}}$ & 0\% \\ 
			$1.000 \times 10^{4}$ & BART & $3.194 \times 10^{-1}$ & $2.167 \times 10^{-2}$ & $1.025 \times 10^{-1}$ & 0\% \\ 
			$1.000 \times 10^{4}$ & DML\_spline & $3.373 \times 10^{-1}$ & $1.202 \times 10^{-2}$ & $1.139 \times 10^{-1}$ &  \\ 
			$1.000 \times 10^{4}$ & DML\_BART & $3.255 \times 10^{-1}$ & $2.052 \times 10^{-2}$ & $1.064 \times 10^{-1}$ & 0\% \\ 
			\hline
		\end{tabular}
		\caption{Noisy confounding simulation results}
		\label{tab:sim_noisy}
	\end{adjustwidth}
\end{table}

\subsection{Less noisy confounding}
\begin{gather*}
	S^a, S^b \sim_{i.i.d.} \unif(-1, 1)\\
	U \sim N( \sin(2\pi S^a*S^b) + S^a + S^b, 0.1)\\
	X \sim N(U^3, 5)\\
	Y \sim N(5U+X, 1)
\end{gather*}
\clearpage
\begin{table}[ht]
	\begin{adjustwidth}{-3cm}{-3cm}
		\centering
		\begin{tabular}{rlrrrr}
			\hline
			n & method & bias & sd & mse & coverage \\ 
			\hline
			$1.000 \times 10^{3}$ & RSR & $4.338 \times 10^{-1}$ & $2.410 \times 10^{-2}$ & $1.888 \times 10^{-1}$ & 0\% \\ 
			$1.000 \times 10^{3}$ & spline & $1.421 \times 10^{-2}$ & $7.492 \times 10^{-3}$ & $2.581 \times 10^{-4}$ & 49\% \\ 
			$1.000 \times 10^{3}$ & gp & $1.464 \times 10^{-2}$ & $7.791 \times 10^{-3}$ & $2.748 \times 10^{-4}$ & 49\% \\ 
			$1.000 \times 10^{3}$ & gSEM & $9.108 \times 10^{-2}$ & $9.040 \times 10^{-3}$ & $8.378 \times 10^{-3}$ & 0\% \\ 
			$1.000 \times 10^{3}$ & spatial+ & $1.768 \times 10^{-2}$ & $7.902 \times 10^{-3}$ & $3.750 \times 10^{-4}$ & 45\% \\ 
			$1.000 \times 10^{3}$ & svc\_mle & $1.401 \times 10^{-2}$ & $7.693 \times 10^{-3}$ & $2.553 \times 10^{-4}$ &  \\ 
			$1.000 \times 10^{3}$ & spline\_interaction & $1.403 \times 10^{-2}$ & $7.515 \times 10^{-3}$ & $2.532 \times 10^{-4}$ & 81\% \\ 
			$1.000 \times 10^{3}$ & gp3 & $8.156 \times 10^{-3}$ & $7.658 \times 10^{-3}$ & $1.251 \times 10^{-4}$ &  \\ 
			$1.000 \times 10^{3}$ & spline3 & $9.351 \times 10^{-2}$ & $6.410 \times 10^{-2}$ & $1.284 \times 10^{-2}$ &  \\ 
			$1.000 \times 10^{3}$ & rf & $-1.034 \times 10^{-1}$ & $2.001 \times 10^{-2}$ & $1.109 \times 10^{-2}$ & 2\% \\ 
			$1.000 \times 10^{3}$ & BART & $1.608 \times 10^{-2}$ & $1.896 \times 10^{-2}$ & $6.173 \times 10^{-4}$ & 100\% \\ 
			$1.000 \times 10^{3}$ & DML\_gp3 & $\mathbf{7.789 \times 10^{-3}}$ & $7.932 \times 10^{-3}$ & $\mathbf{1.235 \times 10^{-4}}$ &  \\ 
			$1.000 \times 10^{3}$ & DML\_spline & $5.611 \times 10^{-2}$ & $5.497 \times 10^{-2}$ & $6.164 \times 10^{-3}$ &  \\ 
			$1.000 \times 10^{3}$ & DML\_BART & $1.143 \times 10^{-2}$ & $1.889 \times 10^{-2}$ & $4.865 \times 10^{-4}$ & 100\% \\ \hline
			$1.000 \times 10^{4}$ & RSR & $4.334 \times 10^{-1}$ & $7.335 \times 10^{-3}$ & $1.879 \times 10^{-1}$ & 0\% \\ 
			$1.000 \times 10^{4}$ & spline & $8.398 \times 10^{-3}$ & $2.300 \times 10^{-3}$ & $7.579 \times 10^{-5}$ & 6\% \\ 
			$1.000 \times 10^{4}$ & gp & $9.312 \times 10^{-3}$ & $2.298 \times 10^{-3}$ & $9.197 \times 10^{-5}$ & 2\% \\ 
			$1.000 \times 10^{4}$ & gSEM & $2.439 \times 10^{-2}$ & $2.461 \times 10^{-3}$ & $6.009 \times 10^{-4}$ & 0\% \\ 
			$1.000 \times 10^{4}$ & spatial+ & $7.380 \times 10^{-3}$ & $2.299 \times 10^{-3}$ & $5.973 \times 10^{-5}$ & 11\% \\ 
			$1.000 \times 10^{4}$ & spline\_interaction & $8.209 \times 10^{-3}$ & $2.306 \times 10^{-3}$ & $7.269 \times 10^{-5}$ & 7\% \\ 
			$1.000 \times 10^{4}$ & spline3 & $4.054 \times 10^{-2}$ & $1.026 \times 10^{-2}$ & $1.748 \times 10^{-3}$ &  \\ 
			$1.000 \times 10^{4}$ & rf & $-5.936 \times 10^{-2}$ & $5.370 \times 10^{-3}$ & $3.553 \times 10^{-3}$ & 0\% \\ 
			$1.000 \times 10^{4}$ & BART & $8.915 \times 10^{-3}$ & $3.818 \times 10^{-3}$ & $9.399 \times 10^{-5}$ & 64\% \\ 
			$1.000 \times 10^{4}$ & DML\_spline & $1.824 \times 10^{-2}$ & $1.001 \times 10^{-2}$ & $4.325 \times 10^{-4}$ &  \\ 
			$1.000 \times 10^{4}$ & DML\_BART & $\mathbf{6.480 \times 10^{-3}}$ & $3.778 \times 10^{-3}$ & $\mathbf{5.621 \times 10^{-5}}$ & 84\% \\ 
			\hline
		\end{tabular}
		\caption{Less noisy confounding simulation results}
		\label{tab:sim_lessnoisy}
	\end{adjustwidth}
\end{table}

\subsection{Smooth exposure}\label{sec:smooth}

\begin{gather*}
	S^a, S^b \sim_{i.i.d.} \unif(-1, 1)\\
	U = \sin(2\pi S^a*S^b) + S^a + S^b\\
	X = C^3 + \cos(2\pi S^a*S^b)\\
	Y \sim N(3C+X, 1)
\end{gather*}

\clearpage
\begin{table}[ht]
	\begin{adjustwidth}{-3cm}{-3cm}
		\centering
		\begin{tabular}{rlrrrr}
			\hline
			n & method & bias & sd & mse & coverage \\ 
			\hline
			$1.000 \times 10^{3}$ & RSR & $6.784 \times 10^{-1}$ & $3.030 \times 10^{-2}$ & $4.611 \times 10^{-1}$ & 0\% \\ 
			$1.000 \times 10^{3}$ & spline & $2.763 \times 10^{-1}$ & $2.757 \times 10^{-2}$ & $\mathbf{7.712 \times 10^{-2}}$ & 0\% \\ 
			$1.000 \times 10^{3}$ & gp & $3.755 \times 10^{-1}$ & $2.176 \times 10^{-2}$ & $1.415 \times 10^{-1}$ & 0\% \\ 
			$1.000 \times 10^{3}$ & gSEM & $5.868 \times 10^{-1}$ & $3.014 \times 10^{-1}$ & $4.350 \times 10^{-1}$ & 73\% \\ 
			$1.000 \times 10^{3}$ & spatial+ & $\mathbf{2.448 \times 10^{-1}}$ & $3.007 \times 10^{-1}$ & $1.502 \times 10^{-1}$ & 94\% \\ 
			$1.000 \times 10^{3}$ & svc\_mle & $3.640 \times 10^{-1}$ & $4.203 \times 10^{-2}$ & $1.342 \times 10^{-1}$ &  \\ 
			$1.000 \times 10^{3}$ & spline\_interaction & $3.513 \times 10^{-1}$ & $6.681 \times 10^{-2}$ & $1.279 \times 10^{-1}$ & 99\% \\ 
			$1.000 \times 10^{3}$ & gp3 & $5.565 \times 10^{-1}$ & $8.876 \times 10^{-2}$ & $3.175 \times 10^{-1}$ &  \\ 
			$1.000 \times 10^{3}$ & spline3 & $1.153 \times 10^{0}$ & $1.582 \times 10^{-1}$ & $1.353 \times 10^{0}$ &  \\ 
			$1.000 \times 10^{3}$ & rf & $5.394 \times 10^{-1}$ & $8.793 \times 10^{-2}$ & $2.986 \times 10^{-1}$ & 0\% \\ 
			$1.000 \times 10^{3}$ & BART & $7.024 \times 10^{-1}$ & $1.265 \times 10^{-1}$ & $5.093 \times 10^{-1}$ & 1\% \\ \hline
			$1.000 \times 10^{4}$ & RSR & $6.750 \times 10^{-1}$ & $9.106 \times 10^{-3}$ & $4.557 \times 10^{-1}$ & 0\% \\ 
			$1.000 \times 10^{4}$ & spline & $2.267 \times 10^{-1}$ & $1.188 \times 10^{-2}$ & $5.152 \times 10^{-2}$ & 0\% \\ 
			$1.000 \times 10^{4}$ & gp & $3.229 \times 10^{-1}$ & $7.812 \times 10^{-3}$ & $1.043 \times 10^{-1}$ & 0\% \\ 
			$1.000 \times 10^{4}$ & gSEM & $2.386 \times 10^{-1}$ & $7.345 \times 10^{-2}$ & $6.231 \times 10^{-2}$ & 10\% \\ 
			$1.000 \times 10^{4}$ & spatial+ & $\mathbf{2.049 \times 10^{-1}}$ & $7.285 \times 10^{-2}$ & $\mathbf{4.726 \times 10^{-2}}$ & 20\% \\ 
			$1.000 \times 10^{4}$ & spline\_interaction & $3.150 \times 10^{-1}$ & $4.915 \times 10^{-2}$ & $1.016 \times 10^{-1}$ & 98\% \\ 
			$1.000 \times 10^{4}$ & spline3 & $1.292 \times 10^{0}$ & $9.115 \times 10^{-2}$ & $1.676 \times 10^{0}$ &  \\ 
			$1.000 \times 10^{4}$ & rf & $5.344 \times 10^{-1}$ & $3.208 \times 10^{-2}$ & $2.867 \times 10^{-1}$ & 0\% \\ 
			$1.000 \times 10^{4}$ & BART & $5.844 \times 10^{-1}$ & $6.789 \times 10^{-2}$ & $3.461 \times 10^{-1}$ & 0\% \\ 
			\hline
		\end{tabular}
		\caption{Smooth exposure simulation results}
		\label{tab:sim_smooth}
	\end{adjustwidth}
\end{table}

\paragraph*{Summary of simulations}

These simulations confirm multiple strands of argument from the body of the paper. First, effect heterogeneity and non-linearity can result in severe bias in models which are unable to account for them. On the other hand, with the noted exception of random forest, flexible models perform well, even competitive with simpler models when the simpler models are correctly specified. DML methods tend to reduce bias and MSE compared to their (single) machine learning counterparts. The assumptions of smooth confounding through space and extra-spatial variation in the exposure of interest have proven to be critical, though small departures from smoothness of the confounding surface were not severely harmful.

\bibliography{bib_protect}       %

\end{document}